\documentclass[aps,prb,twocolumn,superscriptaddress,longbibliography,10pt]{revtex4-2}
\usepackage{graphicx}
\usepackage{array}
\usepackage{amsmath}
\usepackage[dvipsnames]{xcolor}
\usepackage[colorlinks=true, citecolor={blue!80!black}, urlcolor={blue!50!black}, linkcolor = {blue!80!black}]{hyperref}
\usepackage{amssymb}
\usepackage{bm}
\usepackage{color}
\usepackage{bookmark}
\usepackage{tabularx}
\usepackage{mathtools}
\usepackage{microtype}
\usepackage{hhline}
\usepackage{siunitx}

\begin{document}

\title{Full parity phase diagram of a proximitized nanowire island}
\author{J. Shen}
\email{shenjie@iphy.ac.cn}
\affiliation{QuTech and Kavli Institute of Nanoscience, Delft University of Technology, 2600 GA Delft, The Netherlands}
\affiliation{Beijing National Laboratory for Condensed Matter Physics, Institute of Physics, Chinese Academy of Sciences, Beijing 100190, China}
\author{G.W. Winkler}
\affiliation{Microsoft Quantum, Microsoft Station Q, University of California Santa Barbara, Santa Barbara, CA 93106, USA}
\author{F. Borsoi}
\affiliation{QuTech and Kavli Institute of Nanoscience, Delft University of Technology, 2600 GA Delft, The Netherlands}
\author{S. Heedt}
\affiliation{QuTech and Kavli Institute of Nanoscience, Delft University of Technology, 2600 GA Delft, The Netherlands}
\affiliation{Microsoft Quantum Lab Delft, 2600 GA Delft, The Netherlands}
\author{V. Levajac}
\affiliation{QuTech and Kavli Institute of Nanoscience, Delft University of Technology, 2600 GA Delft, The Netherlands}
\author{J.-Y. Wang}
\affiliation{QuTech and Kavli Institute of Nanoscience, Delft University of Technology, 2600 GA Delft, The Netherlands}
\author{D. van Driel}
\affiliation{QuTech and Kavli Institute of Nanoscience, Delft University of Technology, 2600 GA Delft, The Netherlands}
\author{D. Bouman}
\affiliation{QuTech and Kavli Institute of Nanoscience, Delft University of Technology, 2600 GA Delft, The Netherlands}
\author{S. Gazibegovic}
\affiliation{Department of Applied Physics, Eindhoven University of Technology, 5600 MB
Eindhoven, The Netherlands}
\author{R.L.M. Op Het Veld}
\affiliation{Department of Applied Physics, Eindhoven University of Technology, 5600 MB
Eindhoven, The Netherlands}
\author{D. Car}
\affiliation{Department of Applied Physics, Eindhoven University of Technology, 5600 MB
Eindhoven, The Netherlands}
\author{J.A. Logan}
\affiliation{Materials Department, University of California Santa Barbara, Santa Barbara, CA 93106, USA}
\author{M. Pendharkar}
\affiliation{Electrical and Computer Engineering, University of California Santa Barbara, Santa Barbara, CA 93106, USA}
\author{C.J. Palmstr{\o}m}
\affiliation{Materials Department, University of California Santa Barbara, Santa Barbara, CA 93106, USA}
\affiliation{Electrical and Computer Engineering, University of California Santa Barbara, Santa Barbara, CA 93106, USA}
\author{E.P.A.M. Bakkers}
\affiliation{Department of Applied Physics, Eindhoven University of Technology, 5600 MB
Eindhoven, The Netherlands}
\author{L. P. Kouwenhoven}
\affiliation{QuTech and Kavli Institute of Nanoscience, Delft University of Technology, 2600 GA Delft, The Netherlands}
\affiliation{Microsoft Quantum Lab Delft, 2600 GA Delft, The Netherlands}
\author{B. van Heck}
\email{bernard.vanheck@microsoft.com}
\affiliation{Microsoft Quantum Lab Delft, 2600 GA Delft, The Netherlands}

\date{\today}

\begin{abstract}
We measure the charge periodicity of Coulomb blockade conductance oscillations of a hybrid InSb--Al island as a function of gate voltage and parallel magnetic field.
The periodicity changes from $2e$ to $1e$ at a gate-dependent value of the magnetic field, $B^*$, decreasing from a high to a low limit upon increasing the gate voltage.
In the gate voltage region between the two limits, which our numerical simulations indicate to be the most promising for locating Majorana zero modes, we observe correlated oscillations of peak spacings and heights.
For positive gate voltages, the $2e$-$1e$ transition with low $B^*$ is due to the presence of non-topological states whose energy quickly disperses below the charging energy due to the orbital effect of the magnetic field.
Our measurements highlight the importance of a careful exploration of the entire available phase space of a proximitized nanowire as a prerequisite to define future topological qubits.
\end{abstract}

\maketitle

\section{Introduction}

Coulomb blockade conductance oscillations provide quantitative information about the charge and energy spectrum of a mesoscopic island~\cite{vondelft2001}.
The charge periodicity of the oscillations can be directly related to the free energy difference between even and odd fermion parity states of the island \cite{lafarge1993}.
In superconducting islands, the periodicity is $2e$ \cite{lafarge1993,geerligs1990,tuominen1992,eiles1993}, reflecting the presence of a superconducting ground state with even fermion parity.
In gate-defined semiconducting dots, on the other hand, the periodicity is $1e$, up to peak-to-peak variations due the individual energy levels of the dot \cite{johnson1992,foxman1993,alhassid2000}.

\begin{figure}[t!]
\begin{center}
\includegraphics[width=8.6cm]{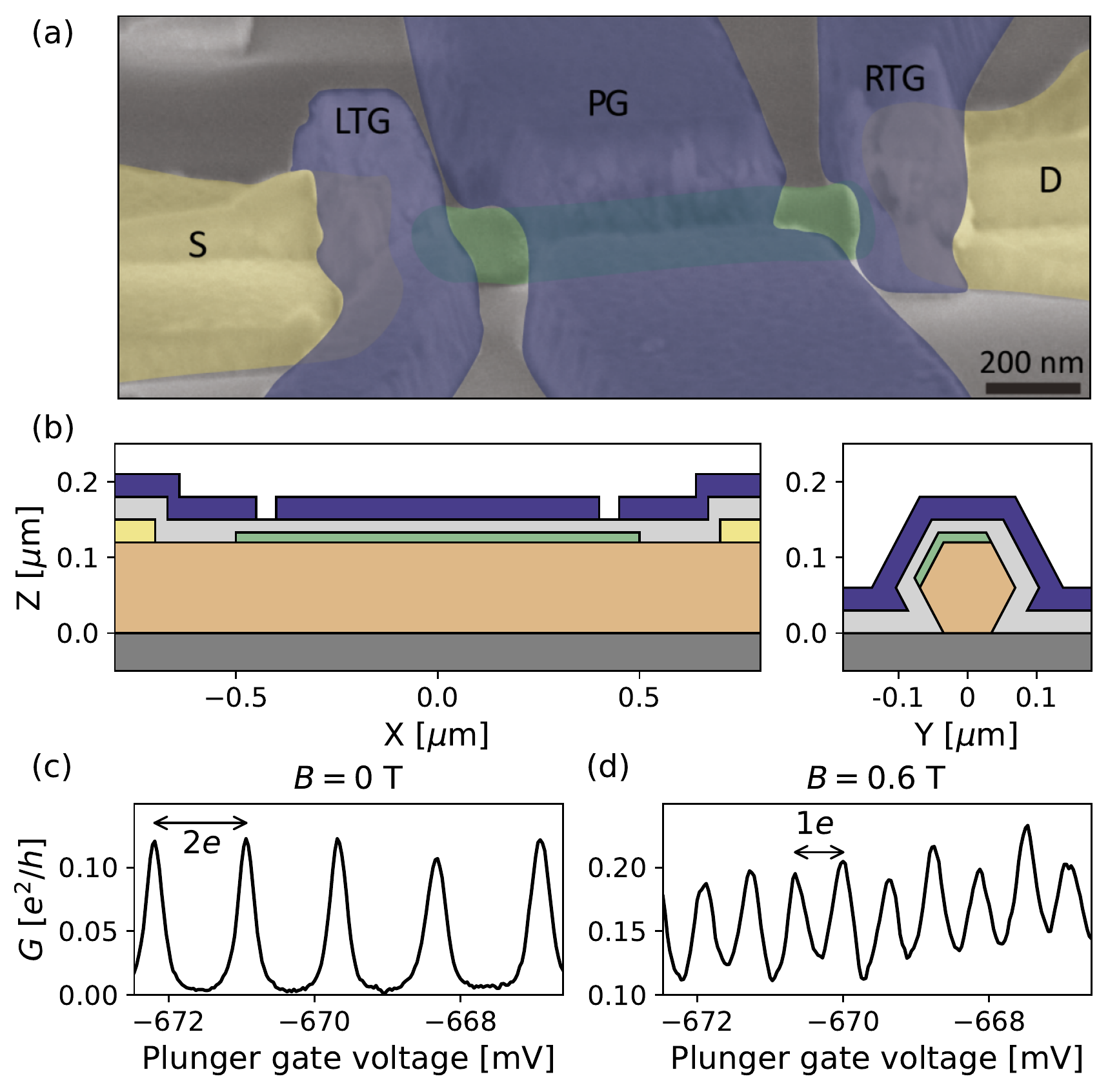}
\caption{\label{fig:1} (a) Scanning electron microscopy image of the experimental device with false colors. Labels indicate source (S), drain (D), left and right tunnel gates (LTG, RTG) and plunger gate (PG). The Al shell is colored in green. (b) Longitudinal (left) and cross-sectional (right) cuts of the model used in the simulations: substrate (dark grey), InSb nanowire (orange), Al (green), ohmic contacts (yellow), dielectric (grey) and gates (blue). Conductance oscillations measured at zero bias voltage exhibit $2e$ peak spacings at $B=0$ (c) and $1e$ peak spacings at $B=0.6$ T (d).}
\end{center}
\end{figure}

Hybrid semiconducting-superconducting islands can be tuned to exhibit both periodicities \cite{higginbotham2015,albrecht2015,albrecht2017,ofarrell2018,shen2018,vaitiekenas2018,vanveen2018,pendharkar2019,vaitiekenas2020,whiticar2020,carrad2020,kanne2020}.
In particular, a magnetic field can be used to tune the periodicity from $2e$ to $1e$, with an intermediate ``even-odd'' regime characterized by a bimodal distribution of peak spacings~\cite{albrecht2015}.
This change in periodicity can be associated with the exciting possibility of a transition into a topological phase with Majorana zero modes~\cite{lutchyn2018,fu2010,heck2016}, with potential applications in topological quantum computing~\cite{nayak2008,karzig2017}.
The $2e$-to-$1e$ transition, however, is a necessary but not sufficient condition to determine the presence of a topological phase~\cite{chiu2017}, since it can be caused by any Andreev bound state~\cite{lee2014,chen2019,prada2020} whose energy decreases below the charging energy of the island.
In fact, early experimental findings on InAs--Al and InSb--Al islands (e.g. Refs.~\cite{albrecht2015,albrecht2017,shen2018,ofarrell2018}) are not fully consistent with a Majorana interpretation.
Possible discrepancies are the decreasing amplitude of even-odd peak spacing oscillations with magnetic field~\cite{dassarma2012,dmytruk2018,cao2018,sharma2020}, as well as the low field at which $1e$-periodicity appeared, compared to the expected value for the topological transition to occur.

In this paper, we report an exhaustive measurement of the Coulomb oscillations in an InSb--Al island as a function of gate voltage and magnetic field.
Our goal is to map out the entire measurable phase space of the island in order to identify potential topological regions and compare their locations to the expected topological phase diagram resulting from state-of-the-art numerical simulations.
We find that the $2e$-to-$1e$ transition happens at a value of the magnetic field, $B^*$, which decreases with increasing gate voltage in agreement with simulations.
Regions with a very low $B^*$ are unlikely to be topological, while the most promising gate range occurs at intermediate values of $B^*$.

\section{Measurement of the parity phase diagram}

The experiment is carried out in the device shown in Fig.~\ref{fig:1}.
It consists of a hybrid InSb--Al nanowire~\cite{gazibegovic2017}, in which two crystallographic facets of the hexagonal InSb cross-section are covered by $8-15$ nm of epitaxial Al film.
The length of the proximitized segment of the nanowire is $\approx\SI{1}{\micro\m}$.
The nanowire is contacted with metallic source and drain leads, and coupled to three gates for electrostatic control.
The two gates on the sides act as tunnel gates,
while the middle gate acts as a plunger gate controlling the electron occupation of the island as well as the cross-sectional profile of the electron density in the semiconductor.
A magnetic field $B$, parallel to the nanowire axis, can be applied to the device.

The device under consideration shows a hard superconducting gap \cite{gazibegovic2017} as well as $2e$-periodic Coulomb oscillations at $B=0$ \cite{shen2018}.
An example of the latter is shown in Fig.~\ref{fig:1}(c), with a $2e$ peak spacing $\approx 1.2$ mV.
From the measurement of the $2e$-periodic Coulomb diamonds ~\cite{shen2020supplementary}, we extract a single-electron charging energy $E_C=e^2/2C\approx \SI{40}{\micro\eV}$ for the island.
In a large magnetic field, the Coulomb oscillations become $1e$-periodic, as shown in Fig.~\ref{fig:1}(d).

\begin{figure}[t!]
\begin{center}
\includegraphics[width=8.6cm]{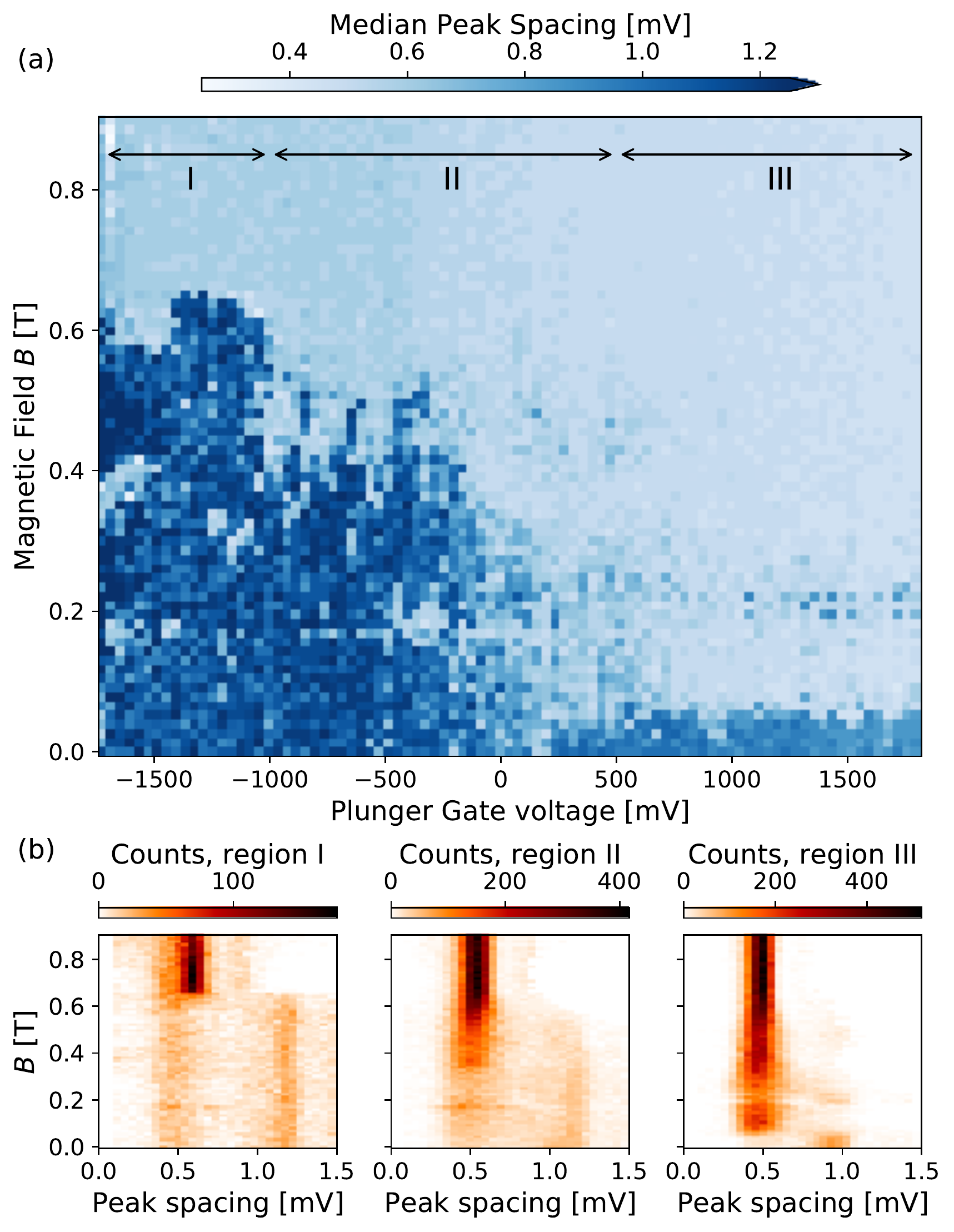}
\caption{\label{fig:2}  (a) Median peak spacing of Coulomb blockade oscillations as a function of magnetic field and gate voltage. Dark blue areas correspond to predominant $2e$ periodicity, light blue to $1e$ periodicity. For each pixel, the median is determined from a window of $20$ mV in plunger gate voltage, corresponding to $\approx 20-40$ conductance oscillations. (b) Peak spacing distributions for regions I, II, and III as labeled in panel (a). We attribute the presence of a residual $1e$ peak at low $B$ in region I to the possible poisoning of the island~\cite{albrecht2017} as well as to the occasional presence of subgap states~\cite{shen2018}.}
\end{center}
\end{figure}

The magnetic field $B^*$ at which the periodicity changes from $2e$ to $1e$ depends on the plunger gate voltage.
To determine this, we have measured a sequence of 90 conductance traces for each magnetic field, centered \SI{40}{\milli\V} apart in the plunger gate and covering a total range of $3.6$ V in the plunger gate as well as \SI{0.9}{T} in magnetic field.
Each trace spans \SI{20}{\milli\V} and contains a sequence of 20 to 40 Coulomb blockade oscillations from which we extract the peak spacings~\cite{shen2020supplementary}.
A telling picture emerges when plotting the median of the peak spacing distribution at each point in parameter space [Fig.~\ref{fig:2}(a)].
This experimental phase diagram can be heuristically divided into three plunger gate voltage regions, which we denote regions I, II, III going from negative to positive gate voltages.

\begin{figure*}[t!]
\begin{center}
\includegraphics[width=17.2cm]{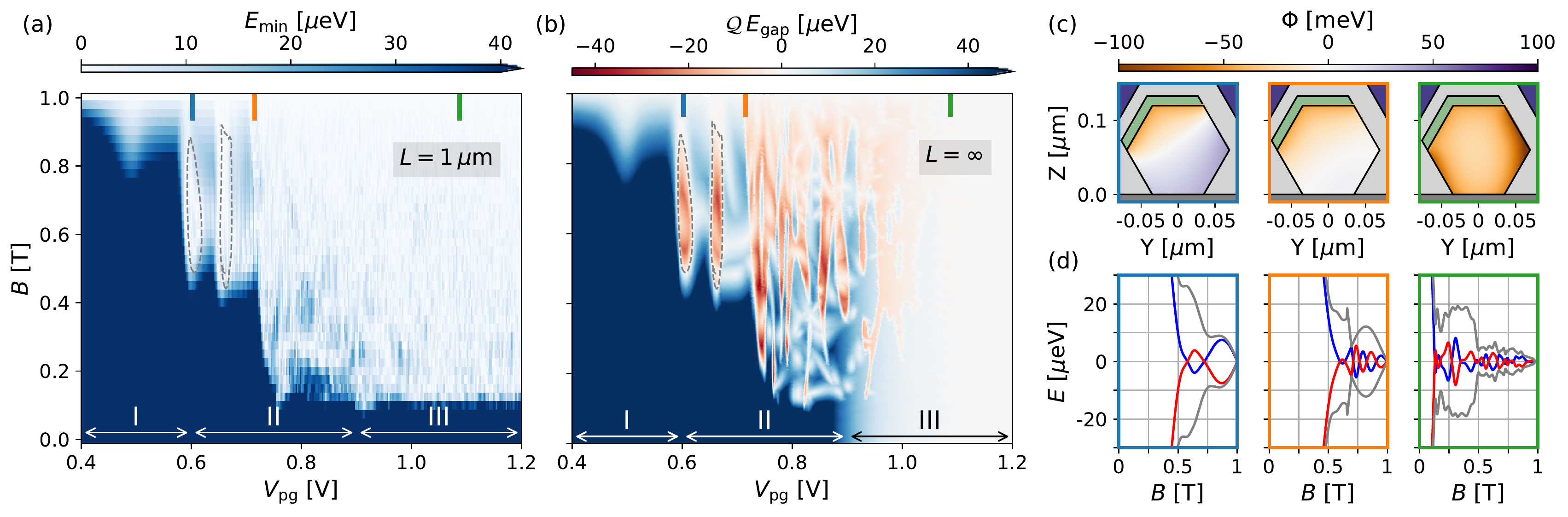}
\caption{(a) Quasiparticle energy gap $E_\textrm{min}$ as a function of plunger voltage $V_\textrm{pg}$ and magnetic field for the simulated island of length~\SI{1}{\micro\m}. We indicate regimes I, II and III as in Fig.~\ref{fig:2}. The energy scale is saturated at $\SI{40}{\micro\eV}$ because this is the estimated charging energy in the experimental device and thus the expected boundary at which the $2e\to1e$ transition would start to occur.
We note that this boundary is only weakly sensitive to the level of disorder used in the simulations, likely due to the short length of the wire~\cite{shen2020supplementary}.
(b) Bulk topological phase diagram indicating the bulk gap $E_\textrm{gap}$ and the sign of the topological index $\mathcal{Q}=\pm 1$, both obtained via the simulation of the bandstructure of an infinitely long wire. Red regions are topological. Two topological regimes with large gap and small coherence length are marked by the dashed grey lines in both panels (a) and (b).
(c) Electrostatic potential profiles in the nanowire cross-section for the three plunger gate values indicated by blue, orange and green bars in panels (a) and (b). 
(d) Magnetic field dependence of the lowest (red/blue) and first excited (gray) energy levels for three different plunger values. The left panel (blue) crosses a topological region of the phase diagram, while the two other panels (orange and green) correspond to topologically trivial regions.
}
\label{fig:3}
\end{center}
\end{figure*}

In region I, the $2e$-to-$1e$ transition occurs at a roughly constant magnetic field $B^*\approx 0.65$ T, slightly lower than the critical field of the Al shell, $B_c\approx 0.8\,$T~\cite{shen2020supplementary}.
This transition is likely caused by quasiparticle poisoning in the superconducting shell, favored by the suppression of pairing in Al~\cite{tuominen1993}.
In region II, $B^*$ decreases gradually with gate voltage, albeit in an irregular fashion.
In region III, $B^*$ is constant and equal to a low value $B^*\approx 50$ mT.
In Fig.~\ref{fig:2}(b) we show the field dependence of the peak spacing distribution for each region.

We note that in Fig.~\ref{fig:2}(a) an even-odd regime, which is present each time the transition from $2e$- to $1e$-periodicity occurs, is likely to be assimilated with the $1e$ regime, because the median does not distinguish a bimodal distribution of spacings from a unimodal one.
The even-odd regime is weakly visible in the standard deviation of the peak spacing distribution, which is larger in the low-field $1e$ regime of regions II and III than in the high-field metallic regime of region I~\cite{shen2020supplementary}.
It is also interesting to notice a weak resurgence of $2e$ spacings at $B\approx 0.2$ T in region III.
Similar results were obtained on another phase diagram measurement~\cite{shen2020supplementary}.

\section{Simulations of a proximitized nanowire island}

To shed light on the parity phase diagram, we perform numerical simulations of a proximitized InSb island.
Advances in the modeling of semiconductor-superconductor hybrid structures allow the inclusion of important effects such as self-consistent electrostatics, orbital magnetic field contribution, and strong coupling between semiconductor and superconductor~\cite{vuik2016, mikkelsen2018, antipov2018, winkler2017, winkler2018, nijholt2016}. 
By integrating out the superconductor into self-energy boundary conditions, we can simulate three-dimensional wires with realistic dimensions including all of the aforementioned effects~\cite{vaitiekenas2020, kringhoj2020}.
This approach takes into account the renormalization of semiconductor properties due to the coupling to the superconductor~\cite{cole2015}.

We model a hexagonal InSb wire with 120~nm facet-to-facet distance and two facets covered by 15~nm Al [Fig~\ref{fig:1}(b)].
In Fig.~\ref{fig:3}(c) we show the simulated electrostatic potential, computed on the level of the Thomas-Fermi approximation~\cite{mikkelsen2018}, inside of the InSb wire for three representative plunger voltages.
Since the concentration of fixed charges in the oxide and interface traps at the oxide-semiconductor interface is not known, the charge environment of the device cannot be determined and the plunger gate values will differ in both range and offset between experiment and numerical simulations, and cannot be quantitatively compared. 
Consistent with the large induced gap observed in InSb--Al devices~\cite{gazibegovic2017}, we assume an electron accumulation layer at the InSb--Al interface~\cite{antipov2018,mikkelsen2018, winkler2018,demoor2018}, with an offset of 50\,meV between the interface pinning of the conduction band in InSb and the Fermi energy of Al.
This choice is also validated via a numerical comparison with the case of a depletion layer~\cite{shen2020supplementary}.
We cannot exclude the presence of band offset fluctuations in the device, an effect not included in the simulations.
The simulations in Fig.~\ref{fig:3} are for a clean InSb wire and a critical field $B_c=1$\,T for Al.

In Fig.~\ref{fig:3}(a) we show the energy gap $E_\textrm{min}$ of an $L=\SI{1}{\micro\meter}$ InSb--Al wire, while in Fig.~\ref{fig:3}(b) we show the bulk energy gap $E_\textrm{gap}$ computed from the bandstructure of an infinitely long wire, with the cross-sectional electrostatic potential chosen to be identical to that which we find in the middle of the $\SI{1}{\micro\meter}$ island.
These simulations identify qualitatively the three plunger gate voltage regions of Fig.~\ref{fig:2} with different regimes of the proximity effect.
The three regimes occur depending on the ratio between the parent superconductor gap $\Delta_\textrm{Al}$, and the semiconductor-superconductor coupling $\Gamma$~\cite{stanescu2011}, which depends on the gate voltage~\cite{antipov2018,winkler2018,demoor2018}.

In region I, $\Gamma\gg \Delta_\textrm{Al}$: InSb is strongly proximitized by Al, leading to significant $g$-factor renormalization such that the induced gap only vanishes when $B$ is close to $B_{c,\mathrm{Al}}$.
This explains the large experimental value of $B^*$ in this region.
The simulations do not include pair-breaking effects in the Al shell, which in reality lead to a regime of gapless superconductivity at $B$ slightly lower than $B_c$~\cite{larkin1965}.
Region II is a crossover region, $\Gamma\approx\Delta_\textrm{Al}$, in which $\Gamma$ and the strength of induced superconductivity gradually decrease with gate voltage.
In region III, $\Gamma$ vanishes for some semiconductor states due to accumulation away from the Al interface~\cite{winkler2018}, and thus the bandstructure is gapless already at $B=0$ [Fig.~\ref{fig:3}(b)].
In this region, the finite wire is \emph{not} gapless [Fig.~\ref{fig:3}(a)]: $E_\textrm{min}$ reaches zero only at a small but finite $B$, similar to what is observed in the experiment.

This surprising feature is a result of finite-size and orbital effects.
In the finite length island, scattering due to the inhomogeneous electrostatic potential at the ends of the wire couples unproximitzed modes and proximitized ones, such that all semiconducting states become gapped~\cite{haim2019,laeven2020}.
Thus, in region III the gap at $B=0$ is finite in Fig.~\ref{fig:3}(a), but not in the bandstructure calculation of Fig.~\ref{fig:3}(b).
However, this gap is fragile: the orbital effect of the magnetic field is strong~\cite{winkler2017} and leads to the gap closing once half of a flux quantum threads the cross-section area $A$, so that $B^*_{\mathrm{III}}\approx h/(4eA)\approx0.1$ T~\cite{winkler2018}.
A comparison with a simulation in which orbital effects are absent~\cite{shen2020supplementary} confirms that they are crucial to explain the data.

\section{Coulomb oscillations in region II}

For inducing topological superconductivity with well-separated Majorana zero modes, region III is unsuitable due to the vanishing bulk gap.
Region II is more promising: in the infinite length limit, it hosts topological phases with a sizable gap, as indicated by the dashed gray lines in Fig.~\ref{fig:3}(b).
In a finite island, identifying these topological phases is hard due to the energy splitting between Majorana zero modes~\cite{dassarma2012,sharma2020}, a problem exacerbated by the narrowness of the topological phases in the plunger gate.
Numerical simulations indicate that the shortest coherence length achievable in the topological phase is $\approx200$~nm, but it occurs only in small pockets of the phase diagram~\cite{shen2020supplementary}.
Even this optimal value leads to a sizable splitting with characteristic field oscillations of increasing amplitude [Fig.~\ref{fig:3}(d)].
To complicate the matter further, similar oscillations can also be observed in topologically trivial regions, as also shown in Fig.~\ref{fig:3}(d).
We note that in our simulations the oscillation amplitude increases with field in the topological phase, but not necessarily in the trivial phase~\cite{dassarma2012,chiu2017,sharma2020}.

\begin{figure}
\begin{center}
\includegraphics[width=8.6cm]{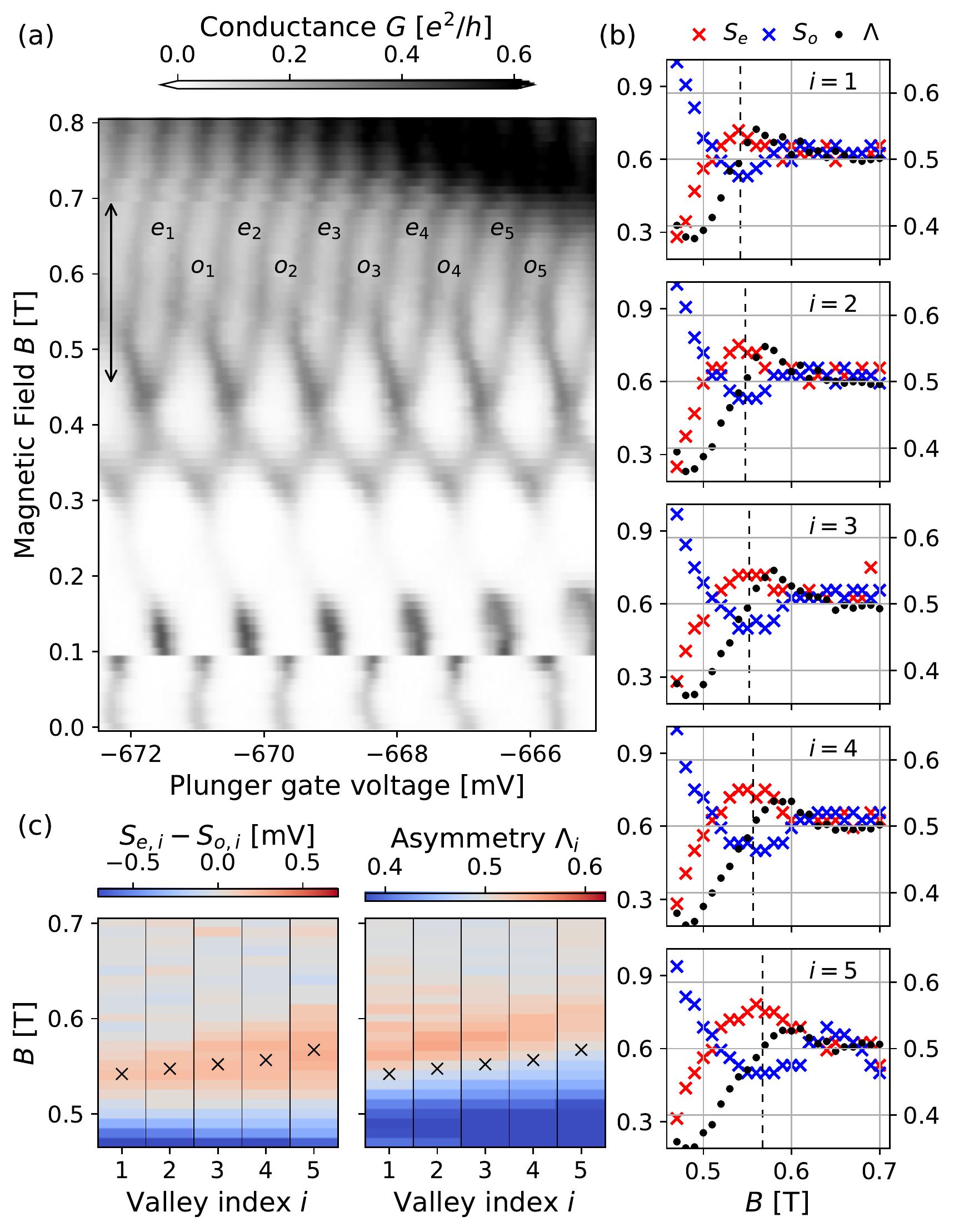}
\caption{\label{fig:4} \emph{(a)} Coulomb blockade oscillations measured versus plunger gate and magnetic field in region II of the phase diagram.
The measurement covers five pairs of even-odd Coulomb valleys (labeled by index $i=1,\dots,5$) in the field range indicated by the black arrow.
\emph{(b)} Field dependence of the even and odd peak spacings $S_\textrm{e,o}$ (left $y$ axis, in mV) and of the peak height asymmetry $\Lambda$ (right $y$ axis) for each pair of Coulomb valleys. Vertical dashed lines denote the linearly interpolated values of $B$ at which $\Lambda=0.5$, corresponding to equal peak heights. These values of $B$ closely match extremal points in $S_\textrm{e,o}$.
\emph{(c)} Peak spacing difference and peak height asymmetry as a function of magnetic field.
Black crosses correspond to the values of $B$ denoted by vertical dashed lines in panel (b).
}
\end{center}
\end{figure}

These energy oscillations can be measured in detail as they reveal themselves in the even-odd peak spacings of conductance oscillations~\cite{albrecht2015}.
An example measured in region II is shown in Fig.~\ref{fig:4}(a).
The $2e$-spaced peaks first split at $B\approx 0.3$ T, leading to a brief $2e$ regime with odd valleys~\cite{shen2018} and then to an even-odd regime, for which we show peak spacing oscillations in Fig.~\ref{fig:4}(b).
The peak spacings undergo one oscillation in magnetic field before the onset of regularly spaced $1e$-peaks at $B\approx 0.65$ T, likely due to poisoning in the Al shell.
The amplitude and position of the peak spacing oscillations change across neighboring valleys, increasing with gate voltage and conferring each valley an individual character [Fig.~\ref{fig:4}(c)].
This shift could be attributed to the strong gate lever arm causing a change in the effective chemical potential of the proximitized InSb bands.

Together with peak spacing oscillations, we also observe oscillating peak heights [Fig.~\ref{fig:4}(b)], captured by the asymmetry parameter $\Lambda = G_{\mathrm{e}\to\mathrm{o}}/(G_{\mathrm{e}\to\mathrm{o}} + G_{\mathrm{o}\to\mathrm{e}})$ where $G_{\mathrm{e}\to\mathrm{o}}$ and $G_{\mathrm{o}\to\mathrm{e}}$ are two neighboring peak heights~\footnote{In Fig.~\ref{fig:4}(b) and (c) we plot the average value over two neighboring $\mathrm{o}\to\mathrm{e}$ peaks for each even valley.}.
$\Lambda$ is related to the electron and hole components of the subgap state mediating the transport at the charge degeneracy point.
In a minimal theory of two coupled Majorana zero modes, it is predicted to oscillate in anti-phase with the energy oscillations~\cite{hansen2018}.
Such a correlation between peak spacing and peak heights is visible in Figs.~\ref{fig:4}(b) and 4(c): in each valley, the symmetric peak heights ($\Lambda=0.5$) occurring at $B\approx 0.55$~T have close-to-maximal peak spacings.
Other datasets taken in region II show similar behavior~\cite{shen2020supplementary}. 
However, in the presence of only a single oscillation we cannot take this as conclusive evidence distinguishing Majorana zero modes from subgap states of trivial origin.

\section{Conclusions}

To conclude, our measurements and simulations have brought to light a mechanism behind the $2e$-to-$1e$ transition in proximitized nanowires, distinct from the transition into a topological phase.
As a consequence, we are able to considerably restrict the range of plunger gate voltage compatible with the presence of Majorana zero-energy modes, although finite-size effects prevent us from a conclusive identification.
A strategy to overcome this obstacle is to measure a sequence of parity phase diagrams as in Fig.~\ref{fig:2} for wires of increasing length.
This would require clean wires to meaningfully compare islands of different length and to preserve the topological phase.
Although the mean free path could not be assessed independently in this study, InSb/Al wires have shown convincing signatures of ballistic transport~\cite{gazibegovic2017}.
Finally, given the importance of an extensive search in the parameter space demonstrated in this work, it will be advantageous to speed up the measurement time by adopting faster measurement techniques~\cite{harabula2017measuring,razmadze2019radio}.

The raw data and the data analysis code at the basis of the results presented in this work are available online~\cite{shen2020data}.
Additional data as well as more information on methods, numerical simulations with additional results including disorder, and data analysis are available in the Supplemental Material~\cite{shen2020supplementary}.

\acknowledgements

We acknowledge stimulating discussions with B.~Bauer, R.~Lutchyn, and T.~Laeven. J.S. acknowledges support from National Science Foundation of China under Grant No. 92065203, Chinese Academy of Sciences under Grant No. XDB33000000 and the Synergic Extreme Condition User Facility. This work has been supported by the European Research Council (ERC), the Dutch Organization for Scientific Research (NWO), the Office of Naval Research (ONR), the Laboratory for Physical Sciences, and Microsoft Corporation.

J.S. fabricated the devices with contribution from D.B. in the optimization of the fabrication recipe; S.G., R.L.M.O.H.V., D.C., J.A.L., M.P., C.J.P., and E.P.A.M.B. carried out the growth of materials; J.S. performed the measurements in collaboration with F.B., S.H., and D.vD; J.S., G.W.W., F.B., S.H., V.L., J.-Y.W., L.P.K., and B.vH discussed and interpreted the experimental data; B.vH and J.S. implemented the experimental data analysis with input from G.W.W., F.B., and S.H; G.W.W. ran the numerical simulations and analyzed the simulation results; J.S., G.W.W., F.B., S.H., L.P.K., and B.vH formulated the comparison between experimental and simulation data; J.S., G.W.W., and B.vH wrote the paper considering input from all authors.

\bibliography{references}

\end{document}


\title{A full parity phase diagram of a proximitized nanowire island: Supplementary Material}
\author{J. Shen}
\email{shenjie@iphy.ac.cn}
\affiliation{QuTech and Kavli Institute of Nanoscience, Delft University of Technology, 2600 GA Delft, The Netherlands}
\affiliation{Beijing National Laboratory for Condensed Matter Physics, Institute of Physics, Chinese Academy of Sciences, Beijing 100190, China}
\author{G.W. Winkler}
\affiliation{Microsoft Quantum, Microsoft Station Q, University of California Santa Barbara, Santa Barbara, CA 93106, USA}
\author{F. Borsoi}
\affiliation{QuTech and Kavli Institute of Nanoscience, Delft University of Technology, 2600 GA Delft, The Netherlands}
\author{S. Heedt}
\affiliation{QuTech and Kavli Institute of Nanoscience, Delft University of Technology, 2600 GA Delft, The Netherlands}
\affiliation{Microsoft Quantum Lab Delft, 2600 GA Delft, The Netherlands}
\author{V. Levajac}
\affiliation{QuTech and Kavli Institute of Nanoscience, Delft University of Technology, 2600 GA Delft, The Netherlands}
\author{J.-Y. Wang}
\affiliation{QuTech and Kavli Institute of Nanoscience, Delft University of Technology, 2600 GA Delft, The Netherlands}
\author{D. van Driel}
\affiliation{QuTech and Kavli Institute of Nanoscience, Delft University of Technology, 2600 GA Delft, The Netherlands}
\author{D. Bouman}
\affiliation{QuTech and Kavli Institute of Nanoscience, Delft University of Technology, 2600 GA Delft, The Netherlands}
\author{S. Gazibegovic}
\affiliation{Department of Applied Physics, Eindhoven University of Technology, 5600 MB
Eindhoven, The Netherlands}
\author{R.L.M. Op Het Veld}
\affiliation{Department of Applied Physics, Eindhoven University of Technology, 5600 MB
Eindhoven, The Netherlands}
\author{D. Car}
\affiliation{Department of Applied Physics, Eindhoven University of Technology, 5600 MB
Eindhoven, The Netherlands}
\author{J.A. Logan}
\affiliation{Materials Department, University of California Santa Barbara, Santa Barbara, CA 93106, USA}
\author{M. Pendharkar}
\affiliation{Electrical and Computer Engineering, University of California Santa Barbara, Santa Barbara, CA 93106, USA}
\author{C.J. Palmstr{\o}m}
\affiliation{Materials Department, University of California Santa Barbara, Santa Barbara, CA 93106, USA}
\affiliation{Electrical and Computer Engineering, University of California Santa Barbara, Santa Barbara, CA 93106, USA}
\author{E.P.A.M. Bakkers}
\affiliation{Department of Applied Physics, Eindhoven University of Technology, 5600 MB
Eindhoven, The Netherlands}
\author{L. P. Kouwenhoven}
\affiliation{QuTech and Kavli Institute of Nanoscience, Delft University of Technology, 2600 GA Delft, The Netherlands}
\affiliation{Microsoft Quantum Lab Delft, 2600 GA Delft, The Netherlands}
\author{B. van Heck}
\email{bernard.vanheck@microsoft.com}
\affiliation{Microsoft Quantum Lab Delft, 2600 GA Delft, The Netherlands}

\date{\today}

\maketitle
\onecolumngrid
\makeatletter
\renewcommand{\theequation}{S\arabic{equation}}
\renewcommand{\thefigure}{S\arabic{figure}}

\section{Additional data}

\begin{figure}
\begin{center}
\includegraphics[width=8.6cm]{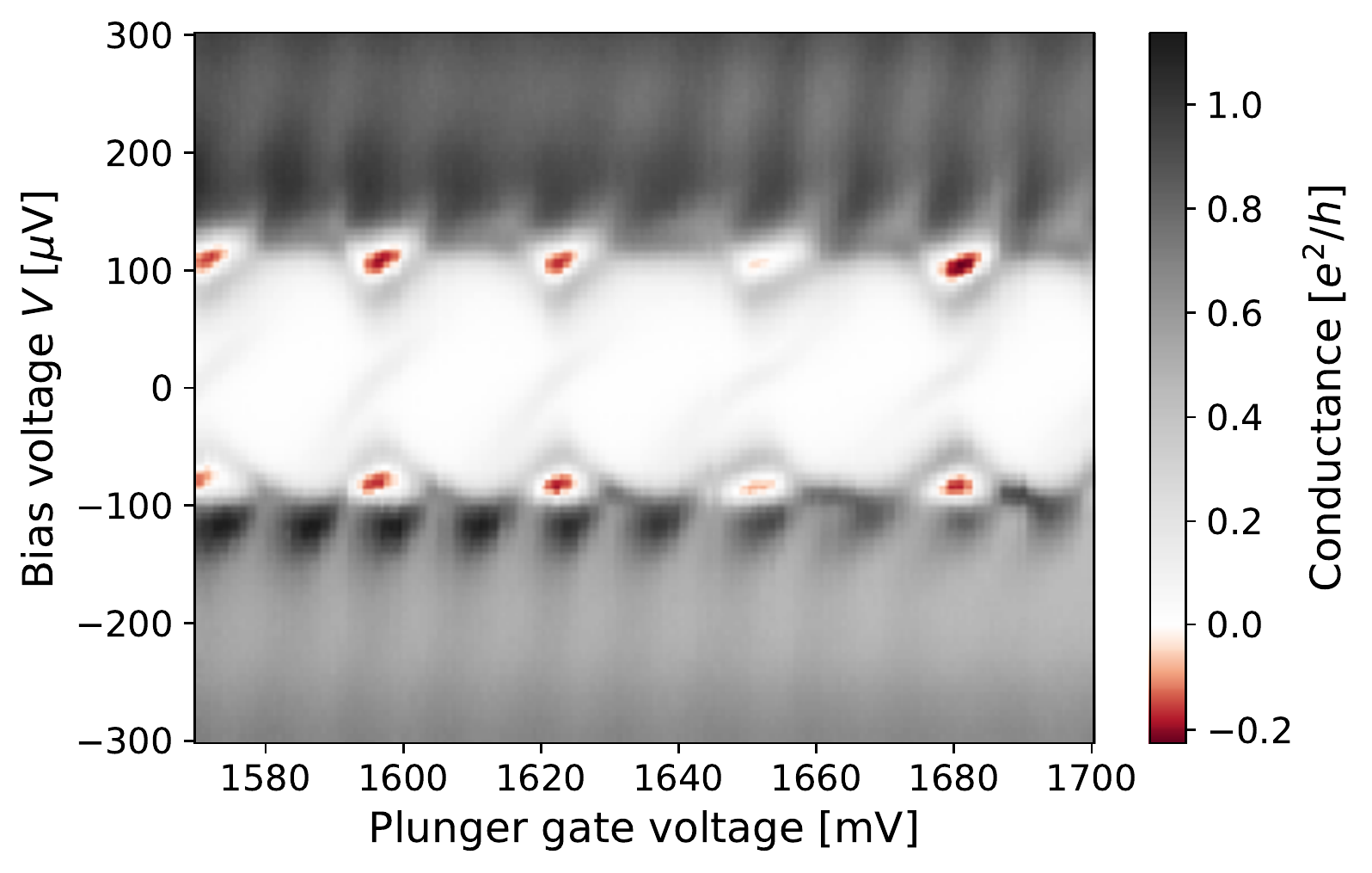}
\caption{Conductance versus bias voltage and plunger gate at $B=0$, showing $2e$-periodic Coulomb oscillations around zero bias voltage. From the height of the Coulomb diamonds~\cite{hergenrother1994}, we estimate a charging energy $E_c\approx\SI{40}{\micro\eV}$.}\label{fig:coulombdiamond}
\end{center}
\end{figure}

\begin{figure}
\begin{center}
\includegraphics[width=8.6cm]{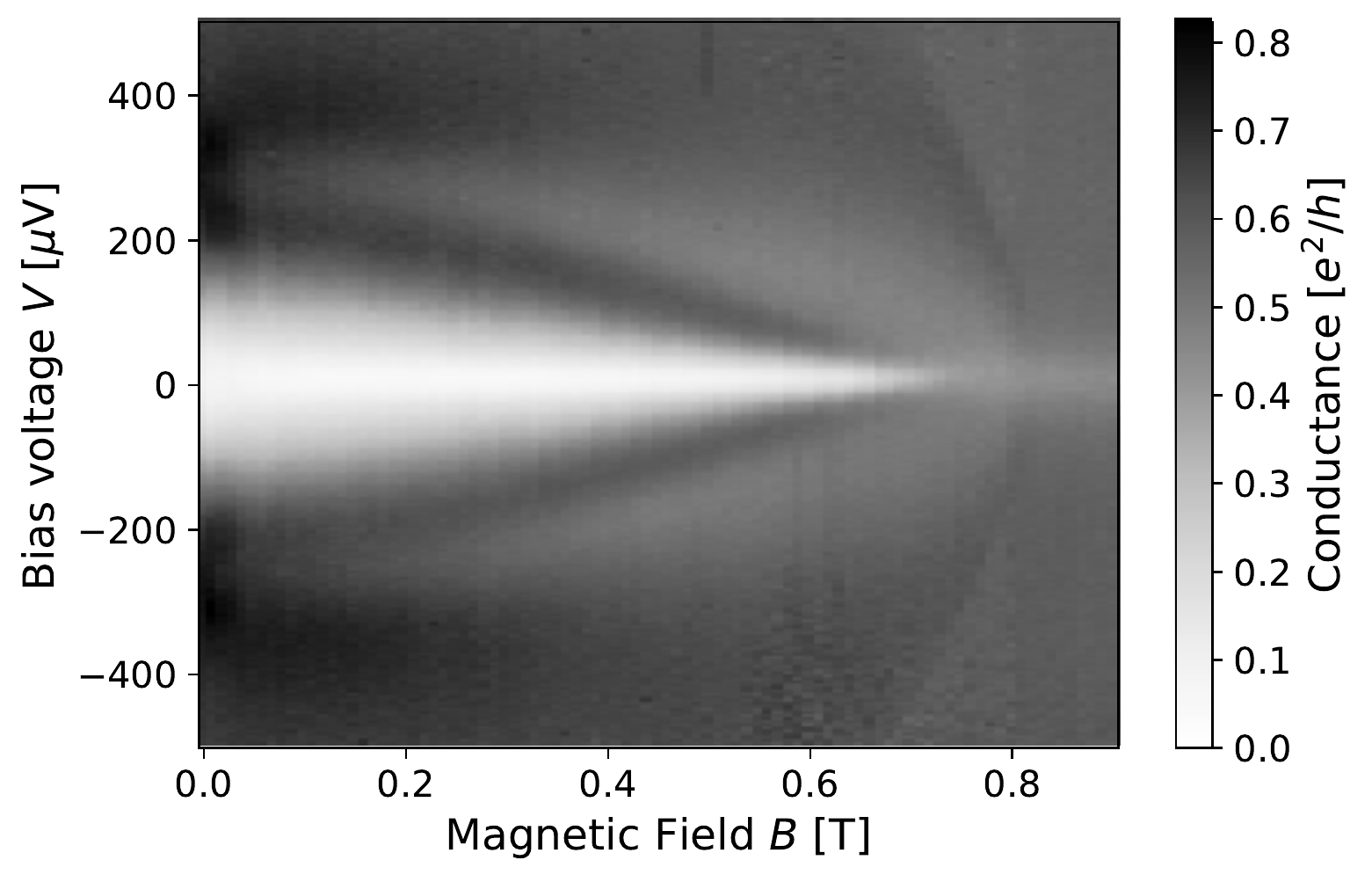}
\caption{Measurement of the parallel critical field of the Al shell of the island via tunneling conductance. This measurement is performed by opening the left tunnel gate ($V_\textrm{LTG}=2$~V), while keeping the right tunnel gate in the tunneling regime ($V_\textrm{RTG}=-2$~V). The critical field is $B_c\approx 0.8$ T.}\label{fig:criticalfield}
\end{center}
\end{figure}

\begin{figure}
\begin{center}
\includegraphics[width=17.2cm]{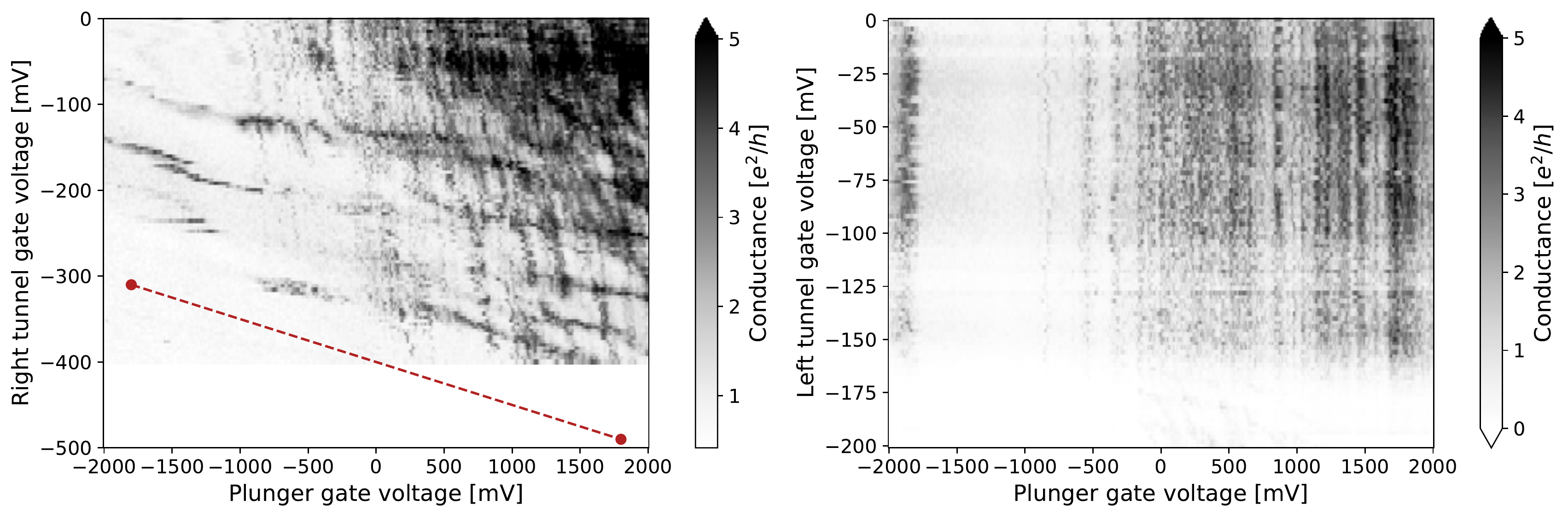}
\caption{Gate versus gate conductance maps, taken at zero bias voltage, illustrating the tunnel gate compensation implemented during the phase diagram measurement of Fig.~2 of the main text. \emph{Left panel:} Right tunnel gate versus plunger gate conductance map, measured at zero bias voltage,  In the phase diagram measurement, the left tunnel gate was kept fixed at $V_\textrm{LTG}=\SI{-0.03}{\V}$, while the right tunnel gate was varied together with the plunger gate following the red dashed line, in order to remain close to the tunneling regime throughout the measurement.~\emph{Right panel:} Same gate map but taken versus the left tunnel gate voltage, in which case no compensation was needed.}\label{fig:gatevsgate}
\end{center}
\end{figure}

\begin{figure}
\begin{center}
\includegraphics[width=17.2cm]{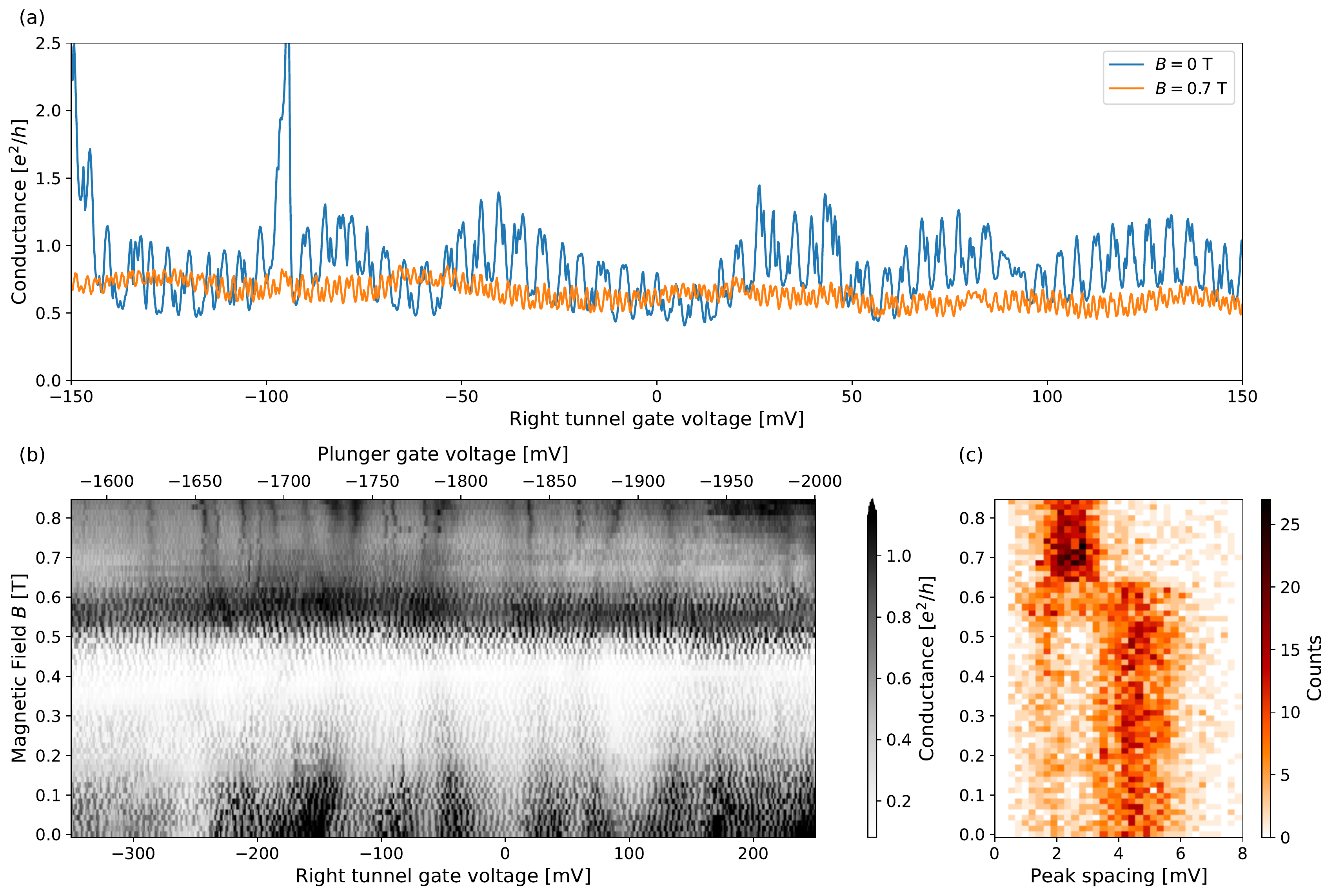}
\caption{\emph{(a-b)} Large-range conductance measurement in the negative plunger gate voltage (region I of the main text). Here the right tunnel gate is varied, with the left tunnel gate kept fixed at $V_\textrm{LTG}=\SI{-0.03}{\V}$. Panel (a) shows a selection of linetraces from panel (b). \emph{(c)} Corresponding peak spacing histogram. The $2e$-$1e$ transition field is not affected by the compensation in tunnel gate voltage and fluctuations in the background conductance.}\label{fig:tgmsmtI}
\end{center}
\end{figure}

\begin{figure}
\begin{center}
\includegraphics[width=17.2cm]{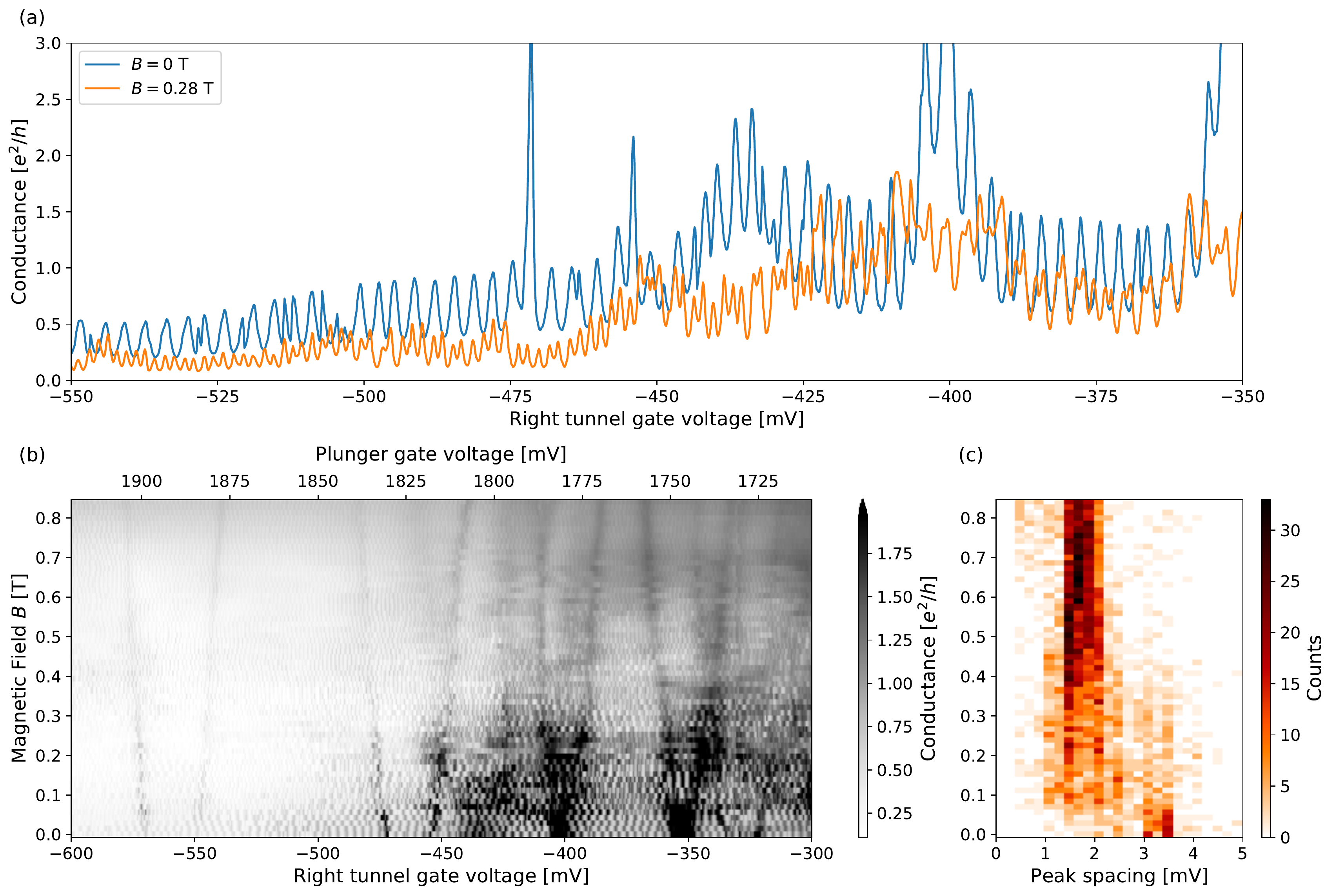}
\caption{\emph{(a-
b)} Large-range conductance measurement in the positive plunger gate voltage (region III of the main text). Here the right tunnel gate is varied, with the left tunnel gate kept fixed at $V_\textrm{LTG}=\SI{-0.03}{\V}$. Panel (a) shows a selection of linetraces from panel (b). \emph{(c)} Corresponding peak spacing histogram. As in the previous Fig.~\ref{fig:tgmsmtI}, the $2e$-$1e$ transition field is not affected by the compensation in tunnel gate voltage and fluctuations in the background conductance.}\label{fig:tgmsmtIII}
\end{center}
\end{figure}

\begin{figure}
\begin{center}
\includegraphics[width=8.6cm]{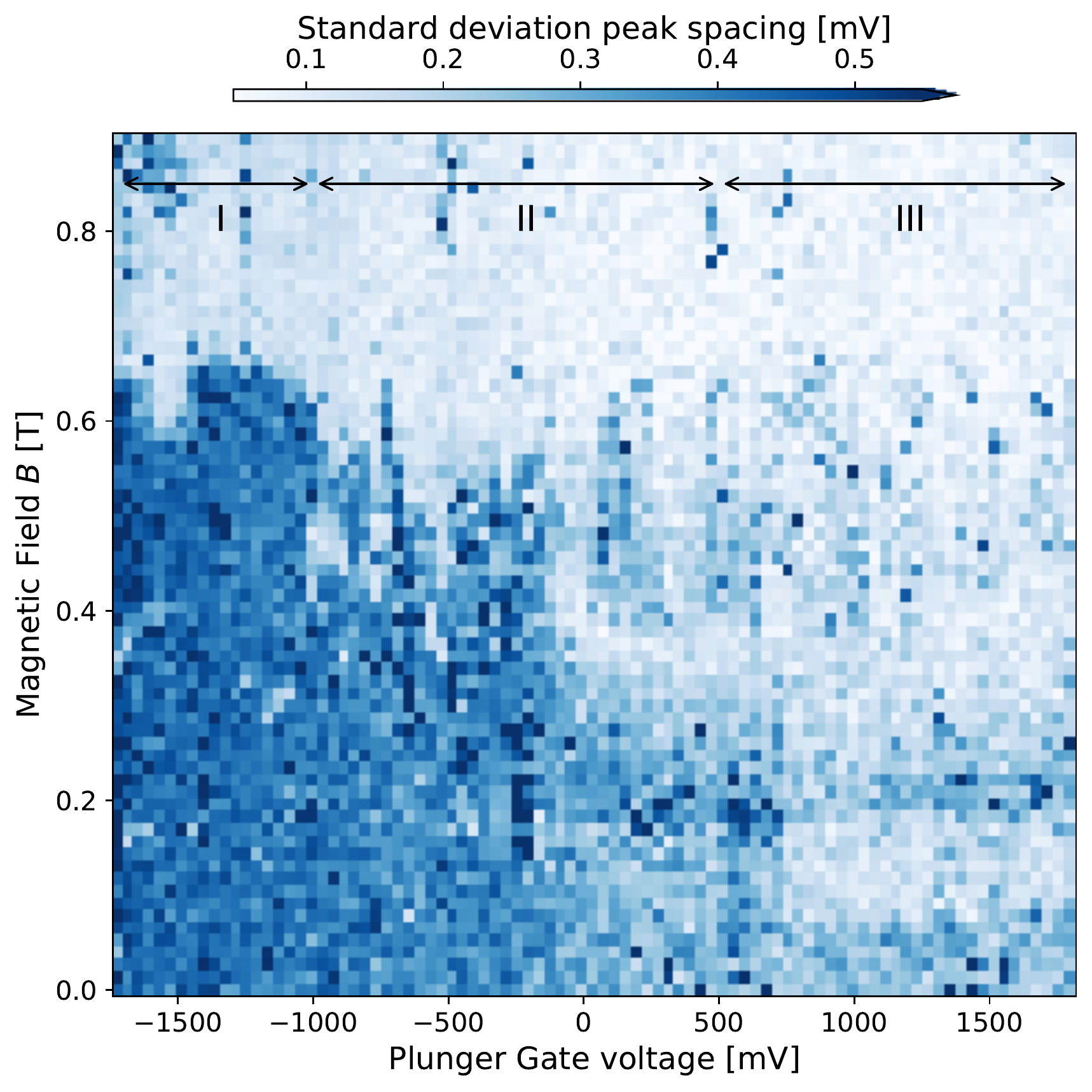}
\caption{Standard deviation of the peak spacings as a function of magnetic field and plunger gate. Regions I, II, III are indicated in analogy to Fig.~2 of the main text. The larger fluctuations occurring at $B\lesssim 0.65$T in regions II and III are an indication of the even-odd regime of Coulomb oscillations, distinguished from the more regular $1e$ regime occurring at $B\gtrsim 0.65$T.}\label{fig:std_spacings}
\end{center}
\end{figure}

\begin{figure}
\begin{center}
\includegraphics[width=8.6cm]{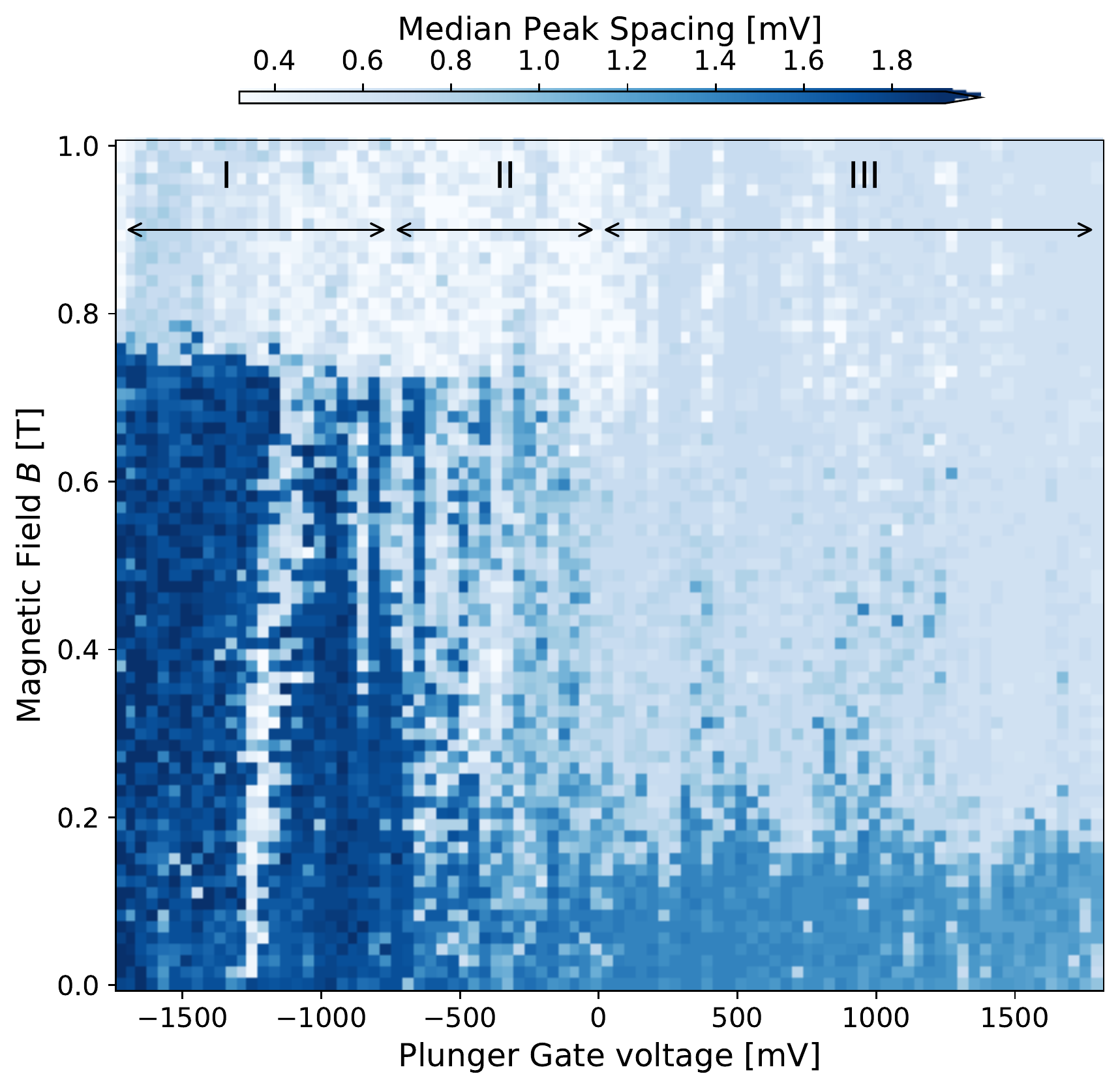}
\caption{A second phase diagram measurement, executed on the same island device measured in Ref.~\cite{shen2018}. This device also shows three characteristic regions as in Fig.~2 of the main text, although the transition between region I and III is more abrupt. This difference may be due to a difference in the lever arm of the gate, or to sample-to-sample as well as local fluctuations in material parameters such as the band offset between InSb and Al. See for instance Fig.~\ref{fig:depletion} for a simulation in the presence of a negative rather than a positive band offset (deplation layer rather than accumulation).
Also, we note the larger value of $B^*$ in region III compared to the phase diagram of Fig.~2 of the main text, which we attribute to the nanowire having a smaller diameter.}\label{fig:second_phase_diagram}
\end{center}
\end{figure}

\begin{figure}
\begin{center}
\includegraphics[width=8.6cm]{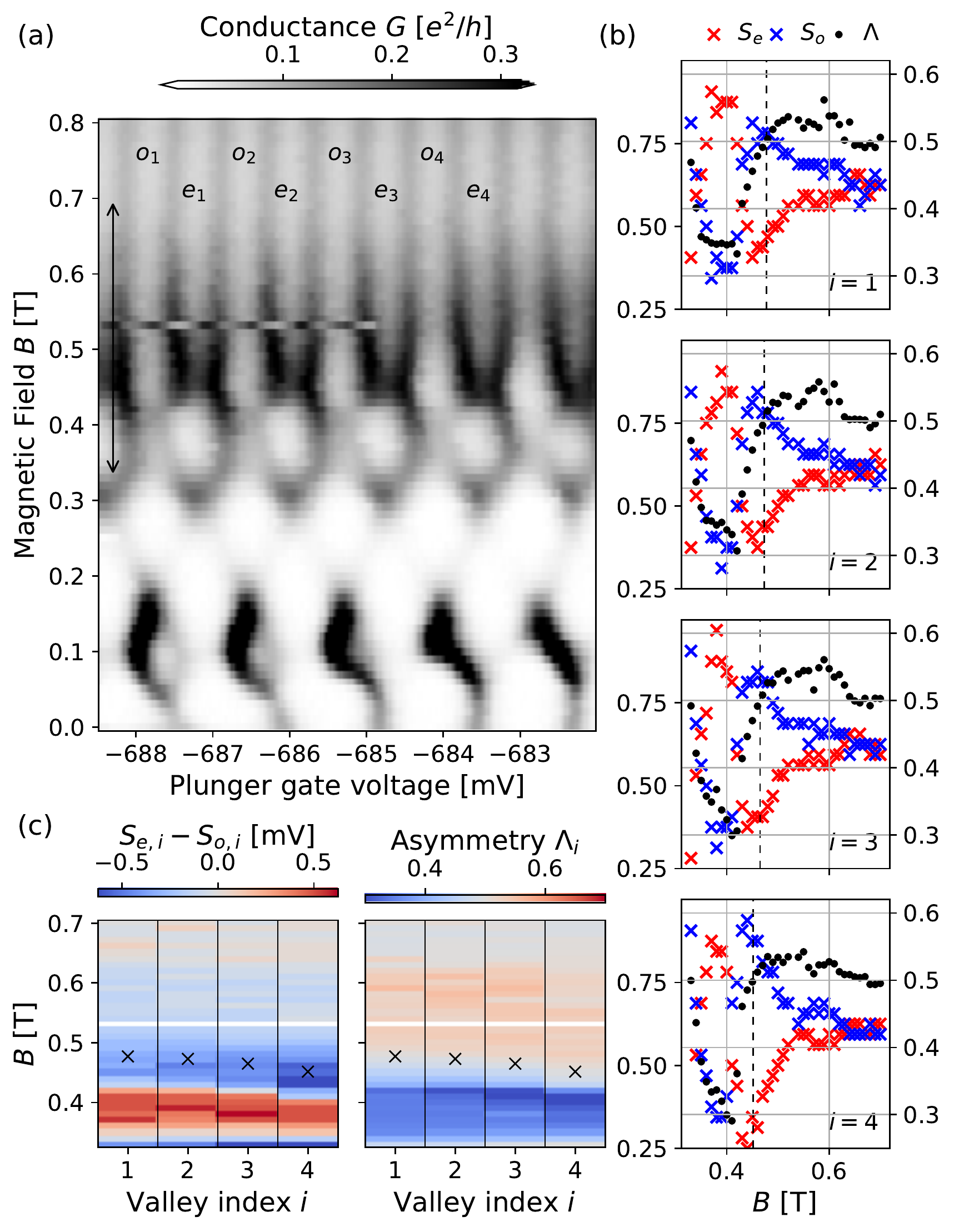}
\caption{Coulomb peak spacings and oscillation analysis, analogous to Fig.~4 of the main text, for an additional dataset measured in region II.
In panels (b) and (c), data points at $B=0.54$~T are omitted since their valley identification is ambiguous due to the gate jump visible in panel (a). Here both tunnel gate voltages are set at \SI{-0.175}{\V}.}\label{fig:S215}
\end{center}
\end{figure}

\begin{figure}
\begin{center}
\includegraphics[width=8.6cm]{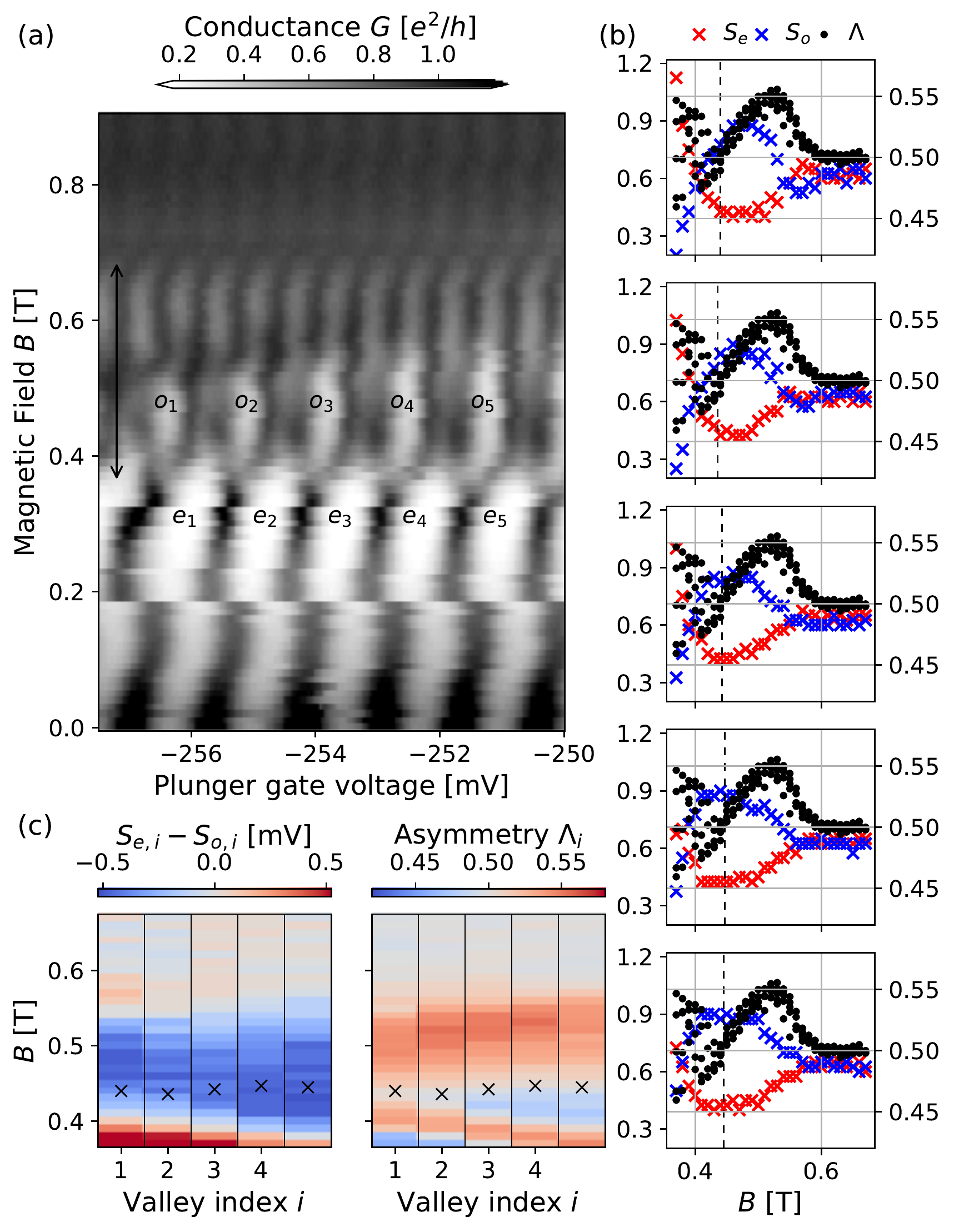}
\caption{Coulomb peak spacings and oscillation analysis, analogous to Fig.~4 of the main text, for an additional dataset measured in region II. Here $V_\textrm{LTG}=$\SI{-0.08}{\V}, $V_\textrm{RTG}=$\SI{-0.3}{\V}.}\label{fig:S170}
\end{center}
\end{figure}

\begin{figure}
\begin{center}
\includegraphics[width=12.9cm]{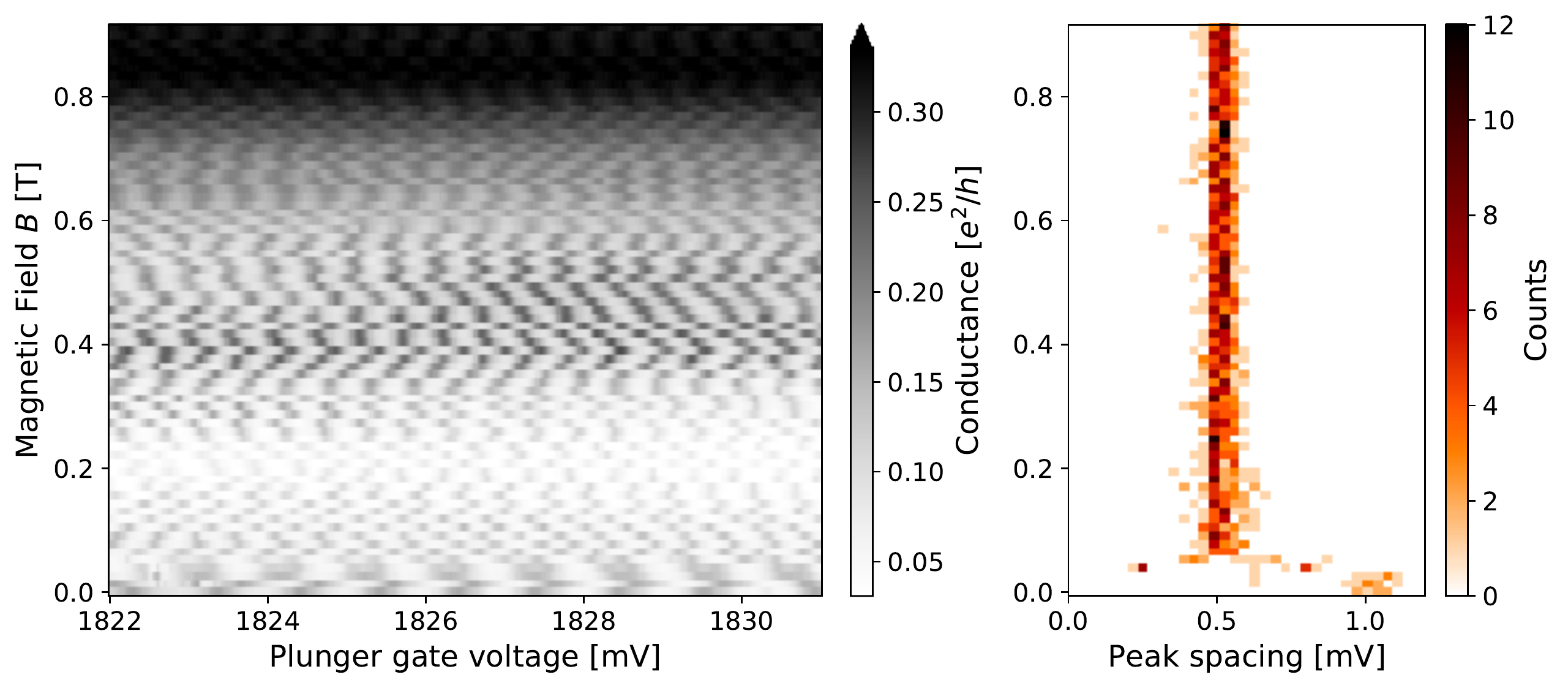}
\caption{Coulomb peak oscillations and peak spacing extraction for an additional dataset in region III. We note that with this resolution in the conductance data, the even-odd regime is briefly visible in the field dependence of the peak spacing histograms in the right panel, around $50$ mT. Peak spacing oscillations are instead not visible at higher fields.}\label{fig:S114}
\end{center}
\end{figure}

\begin{figure}
\begin{center}
\includegraphics[width=12.9cm]{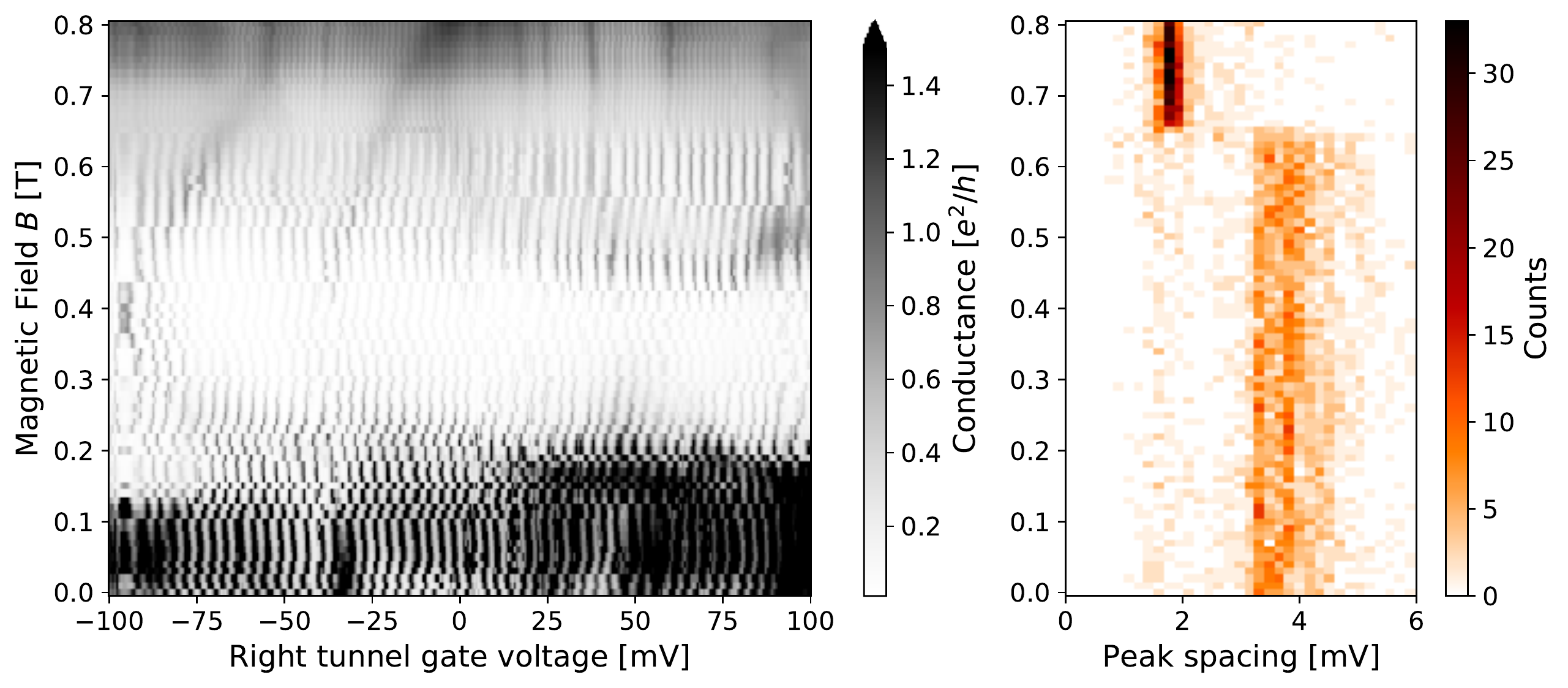}
\caption{Coulomb peak oscillations and peak spacing extraction for an additional dataset in region I, measured as a function of the tunnel gate voltage at a fixed value of the plunger gate voltage, $V_\textrm{PG}=$\SI{-1.4}{\V}.}\label{fig:S267}
\end{center}
\end{figure}

\begin{figure*}
\begin{center}
\includegraphics[width=17.2cm]{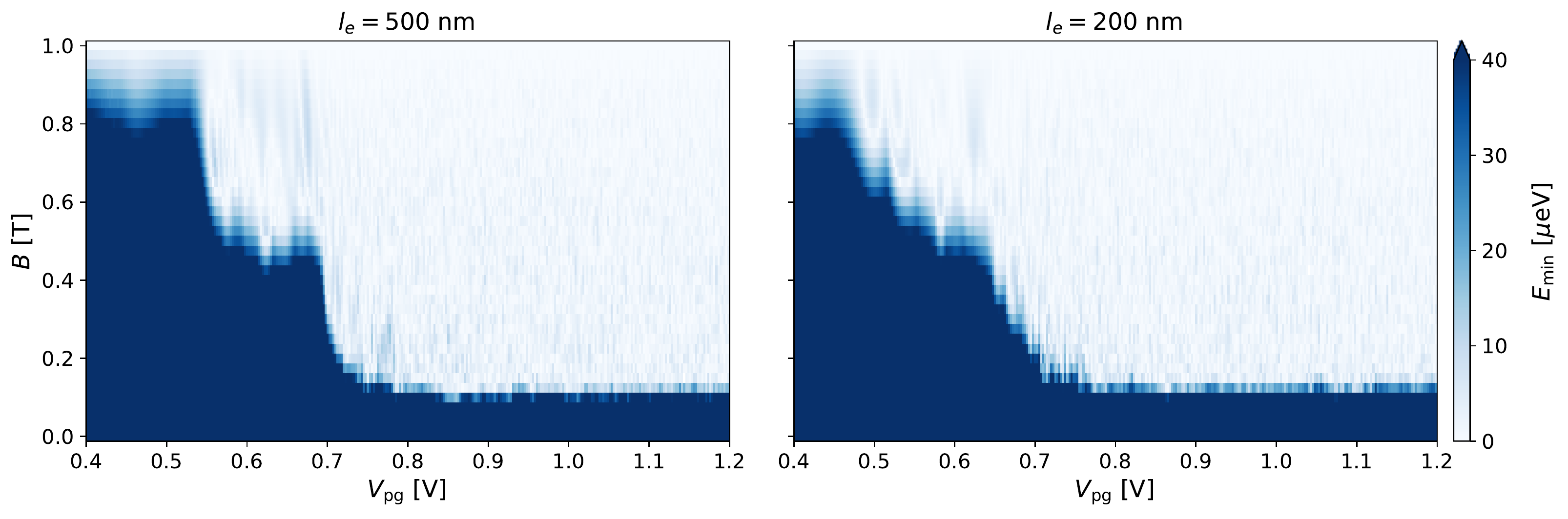}
\caption{Energy gap $E_\textrm{min}$ as a function of plunger voltage and magnetic field in the simulated island with length \SI{1}{\micro\meter}, for two different values of mean free path $l_e$. A single disorder realization is shown. We find that with increased disorder strength, and decreased mean free path, the boundary between the dark blue and white regions of the parameter space, i.e. the boundary at which $E_\textrm{min}$ first vanishes, becomes more and more smeared due to disorder broadening of the individual bands.}\label{fig:mfp}
\end{center}
\end{figure*}

\begin{figure}
\begin{center}
\includegraphics[width=8.6cm]{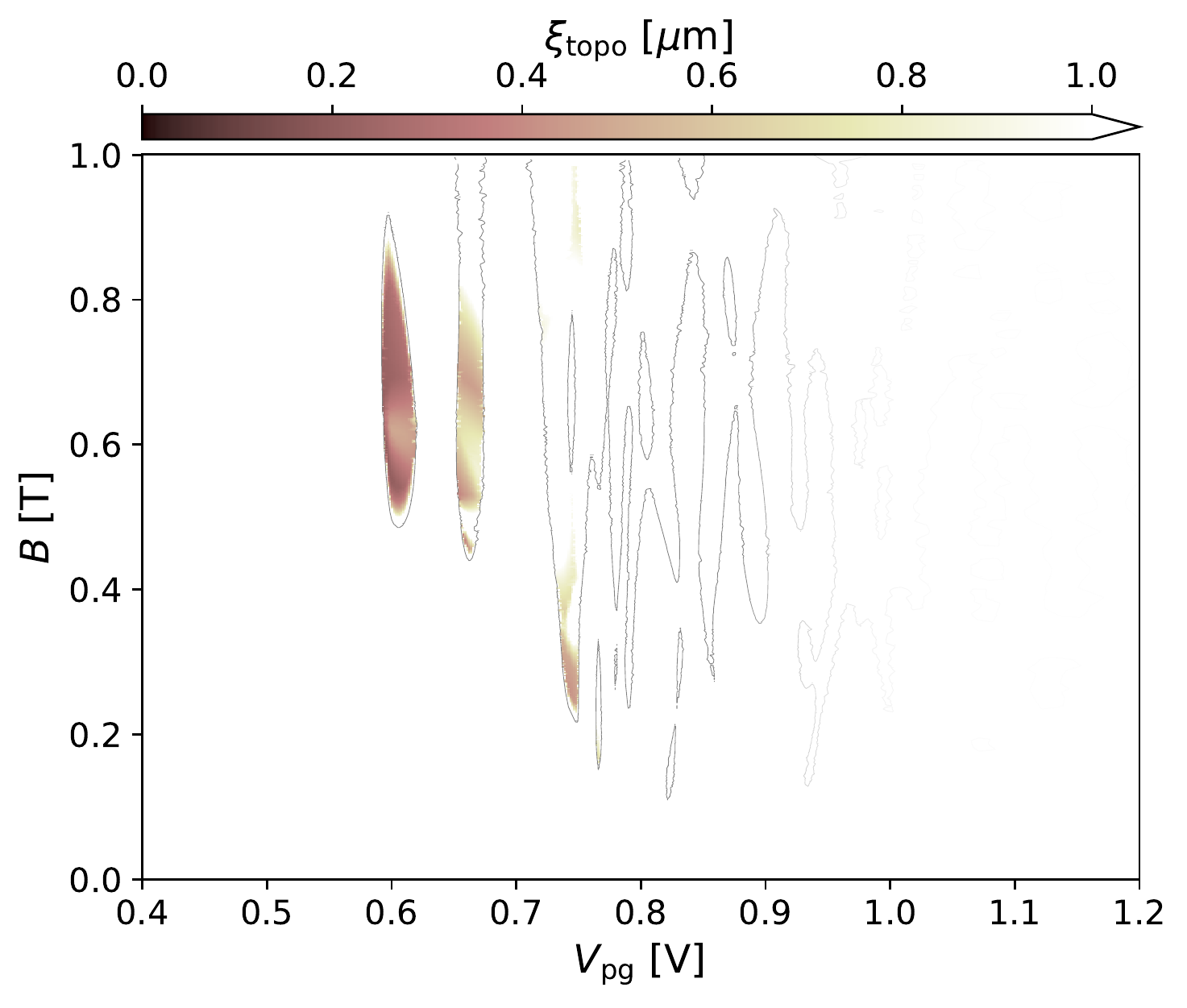}
\caption{Topological coherence length as a function of plunger voltage and magnetic field in the bulk wire. The topological phase boundaries from the middle panel of Fig.~3b of the main text are shown as light grey lines. Only two topological phases at the lowest plunger values have a coherence length that is significantly shorter than the island length.}\label{fig:xi}
\end{center}
\end{figure}

\begin{figure}
\begin{center}
\includegraphics[width=8.6cm]{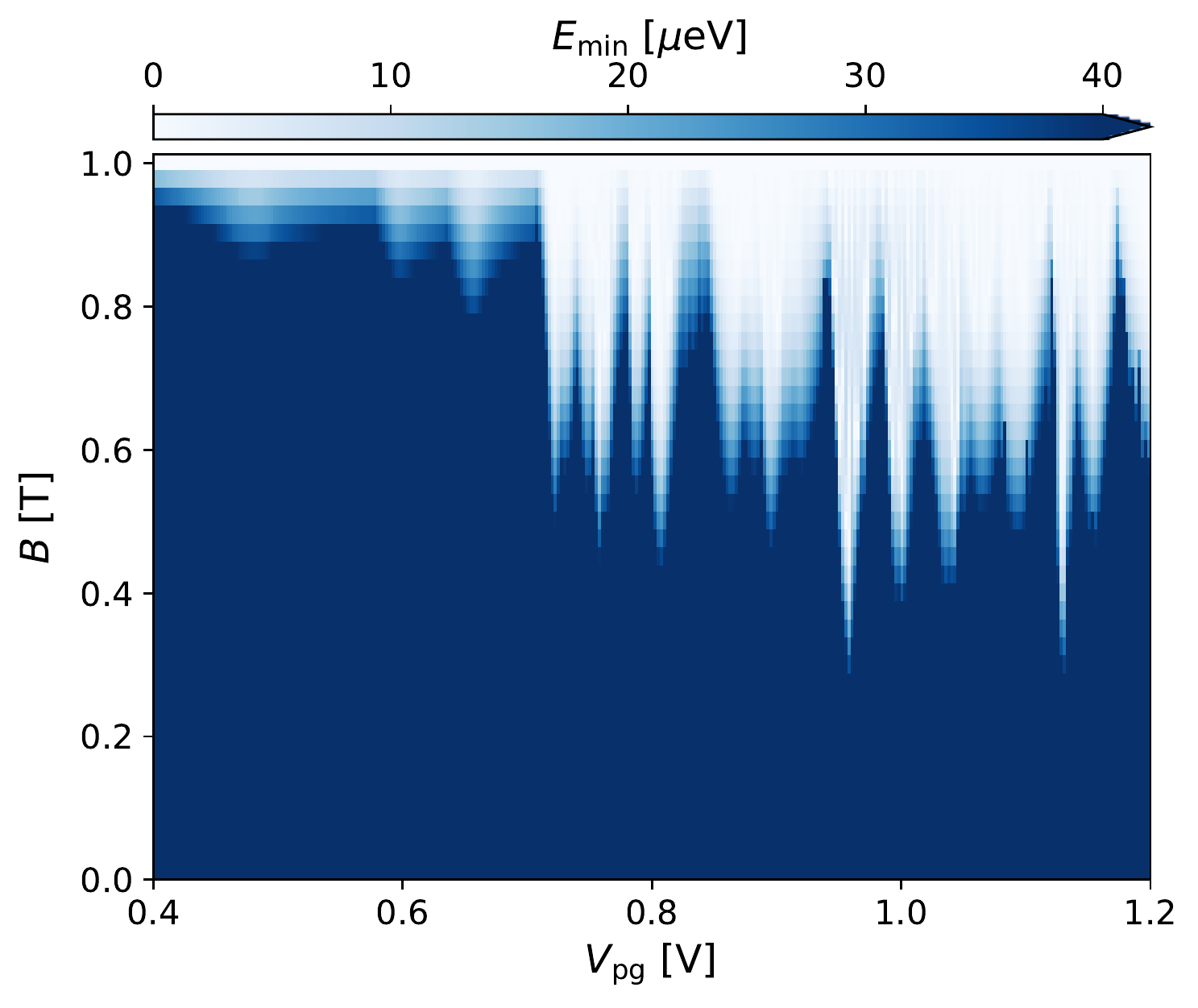}
\caption{Energy gap as a function of plunger voltage and magnetic field in the simulated island with the orbital effect of the magnetic field turned off in the simulation. In comparison to the left panel of Fig.~3 of the main text the energy gap is not suppressed until much larger magnetic fields. Furthermore, the transition is strongly dependent on plunger gate voltage for voltages corresponding to region III in Fig.~3 of the main text.}\label{fig:no_orbital}
\end{center}
\end{figure}

\begin{figure}
\begin{center}
\includegraphics[width=8.6cm]{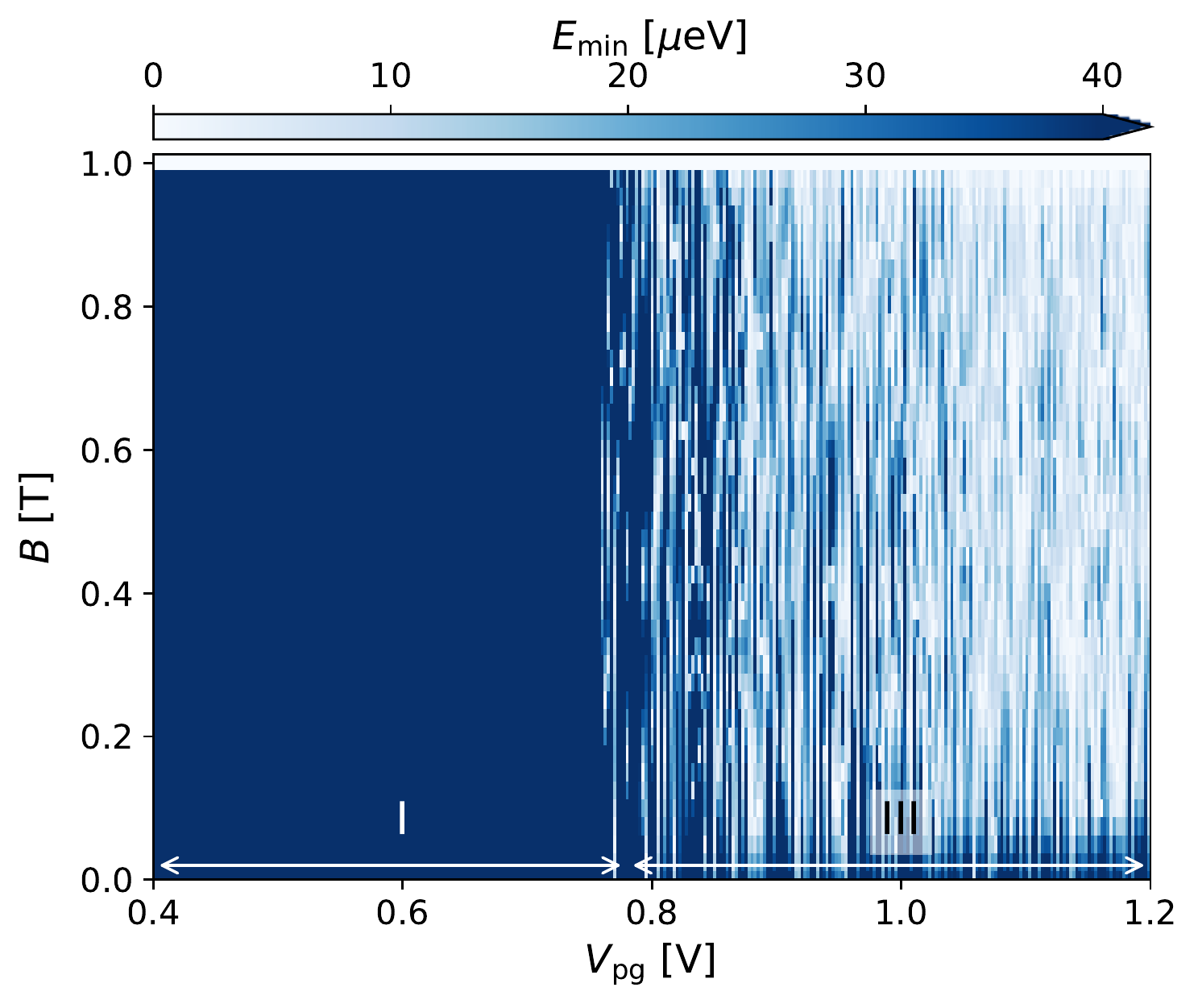}
\caption{Energy gap as a function of plunger voltage and magnetic field in the simulated island with a band offset of -25\,meV. Here the negative band offset corresponds to a depletion layer at the InSb/Al interface. In comparison to Fig.~3a, the transition region II seems very small or absent and region I directly transitions into region III. Due to the absence of a transitional region II, the depletion layer case is not promising for topological applications. We note that spatial fluctuations of the band offset are not captured in the simulation, but they may occur in experimental devices.}\label{fig:depletion}
\end{center}
\end{figure}

\begin{figure}
\begin{center}
\includegraphics[width=17.2cm]{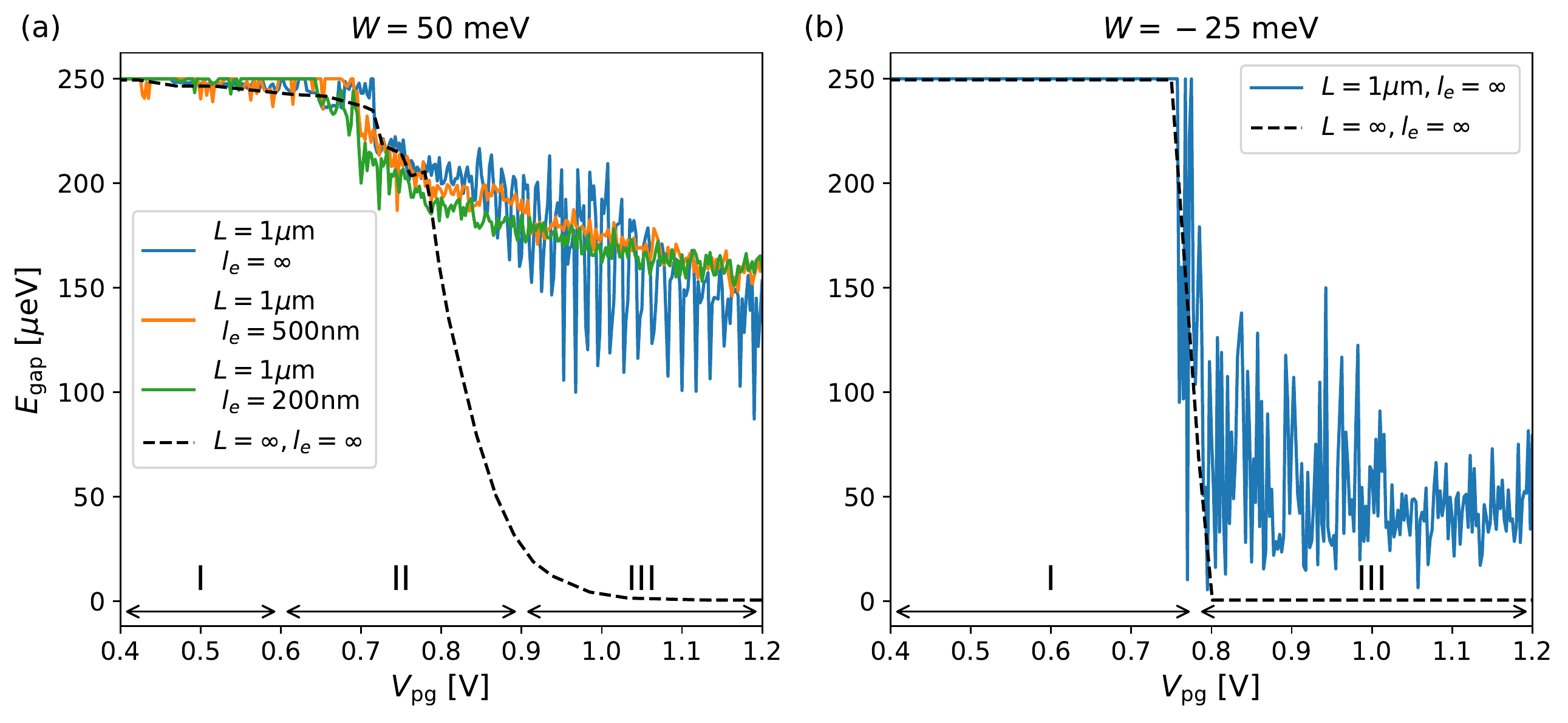}
\caption{Energy gap as a function of plunger voltage at zero magnetic field for a band offset of (a) 50\,meV and (b) -25\,meV. We show the gap both in a clean infinite system, and in a finite system with and without disorder. Both the gap in a finite system with and without disorder and infinitely long system is shown. The dashed blacked line in the left panel most clearly identifies the three regimes of the proximity effect described in the main text: the curve of $E_\textrm{gap}$ versus plunger gate shows two plateaus separated by one crossover region. The first plateau, with $E_\textrm{gap}\approx \Delta_\textrm{Al}$, is identified with region I. The second plateau, with $E_\textrm{gap} = 0$, is identified with region III. The crossover region is region II. Finally, the difference between the energy gap of the finite system and of the infinite system in region III is clearly visible.}\label{fig:gap}
\end{center}
\end{figure}

We include in this Supplementary the following data:

\begin{itemize}
	\item A voltage bias versus plunger gate scan showing $2e$-periodic Coulomb diamonds at $B=0$, Fig.~\ref{fig:coulombdiamond}.
	\item A tunneling conductance measurement showing the parallel critical field of the Al shell, Fig.~\ref{fig:criticalfield}.
	\item A plunger gate versus tunnel gate conductance scan, which is a calibration measurement for the phase diagram measurement reported in the main text, Fig.~\ref{fig:gatevsgate}.
	\item Two additional large-range conductance measurements of Coulomb blockade oscillations as a function of plunger gate and magnetic field, showing in larger detail the $2e$-$1e$ transition occuring at different fields in region I and III, as discussed in the main text; see Fig.~\ref{fig:tgmsmtI} and \ref{fig:tgmsmtIII}.
	\item The standard deviation of the peak spacing distribution measured as a function of plunger gate and magnetic field for the same data used in the phase diagram of the main text, see Fig.~\ref{fig:std_spacings}.
	\item A second parity phase diagram obtained for a different device than considered in the main text, Fig.~\ref{fig:second_phase_diagram}. This second device is the one previously studied in Ref.~\cite{shen2018}.
	\item Two additional datasets taken in region II of the device discussed in the main text, showing Coulomb peak and height oscillations as a function of magnetic field, see Fig.~\ref{fig:S215} and~\ref{fig:S170}.
	\item Two additional datasets taken in region I and III of the device discussed in the main text, showing Coulomb peak oscillations as a function of magnetic field, see Fig.~\ref{fig:S114} and~\ref{fig:S267}. These datasets illustrate the qualitative difference between Coulomb oscillations in the three regions identified in the main text.
	\item In Fig.~\ref{fig:mfp} we show numerical simulations of the energy gap similar to the left panel of Fig.~3 of the main text, but with an uncorrelated disorder potential. 
	\item In Fig.~\ref{fig:xi} we show numerical simulations of the topological coherence length in the bulk wire for the same parameters as the middle panel in Fig.~3 of the main text. 
	\item In Fig.~\ref{fig:no_orbital} we show numerical simulations of the energy gap similar to the left panel of Fig.~3 of the main text, but with the orbital effect of the magnetic field turned off. 
	\item In Fig.~\ref{fig:depletion} we show numerical simulations of the energy gap, similar to Fig.~3a of the main text, but with a depletion instead of an accumulation layer at the InSb/Al interface.
	\item In Fig.~\ref{fig:gap} we show numerical simulations of the energy gap at zero magnetic field for finite and infinite systems with accumulation and depletion layers.
\end{itemize}

We refer to our earlier study of a nominally identical device~\cite{shen2018} for further technical details on growth, fabrication and measurements.

\section{Data analysis}
\label{sec:data_analysis}

Our data analysis routines are fully accessible and reproducible via the published data and code accompanying this manuscript~\cite{shen2020data}. Here, for the convenience of the reader, we describe the main steps required to go from the raw data to the phase diagram reported in the main text.

The raw data underlying the phase diagram consists of a conductance trace taken for each of the plunger gate and magnetic field values reported in the figure.
Each conductance trace consists of 401 data points measured in a fine plunger gate voltage range of 20 mV centered around mean values spaced by 40 mV.
The raw data consists of both lock-in quadratures of the conductance measurements.
The data analysis steps are executed using standard python libraries and routines, and can be summarized as follows:

\begin{itemize}
	\item We rotate the lock-in quadratures so that the real part of the signal is maximized, and the imaginary part only contains noise. This step is necessary as we find that the amplification of the signal in the measurement chain introduces an unwanted rotation in the quadrature plane. See Ref.~\cite{zhang2021large} for a detailed explanation of the origin of this issue. We note that this step does not influence our results because they rely only on the positions of the conductance peaks, which is unaffected by the procedure.
	\item We smoothen each conductance using a Savatzky-Golay polynomial filter, in order to remove noise fluctuations in the data and facilitate the subsequent steps.
	\item We run a peak finder routine available in the SciPy package~\cite{scipy} to determine the positions of conductance peaks. This algorithm simply determines whether a conductance point is a local maximum by comparison with its neighboring points.
	\item The resulting peaks are filtered using a combination of thresholds: a small distance threshold to eliminate double counting of single peaks due to noise, a height threshold to eliminate spurious peaks, and finally a $z$-score height threshold to eliminate remaining outliers. Because the average amplitude of the signal varies of the large dataset, the height thresholds are implemented relative to the average magnitude for each conductance trace. The results of the automated filtering of the peaks was double-checked by eye against the original, unfiltered data in order to settle on reasonable filtering parameters that would remove most peaks originating from noise while keeping the large majority of genuine Coulomb peaks. Boundary cases, e.g. weak Coulomb peaks right above the noise level or repeated peaks due to gate jumps, may be filtered or kept erroneously. We have verified the the resulting phase diagram is robust to specific choices of filtering parameters, provided that they correspond to a decent filtering of the data.
\end{itemize}

Possible sources of noise in the resulting processed data shown in Fig.~2 of the main text are the following.
First, although we compensated the tunnel gates during the plunger gate sweep (see Fig.~\ref{fig:gatevsgate}), the conductance level varied over the measurements, leading to a varying amplitude of the background signal.
Second, the occurrence of uncontrolled resonances under the tunnel gates sometimes leads to conductance oscillations with spurious periodicity.
Third, the lever arm of the plunger gate changes slightly over the entire measurement range, as can be noticed by the shift of the median peak spacing magnitude in the $1e$ regime at high $B$.
Fourth, charge fluctuations in the dielectric led to the occurrence of undetected gate jumps during the long measurement, causing missing or doubly-counted peaks and, thus, irregularities in the resulting peak spacing distributions.
While these factors contribute to fluctuations in the peak spacing distributions extracted in Fig.~2b, in particular in the $2e$ regime of region I, overall they do not obfuscate the emerging structure of the parity phase diagram.

\section{Numerical Simulations}
\label{sec:num}

The numerical simulations are performed with the same code as described in Ref.~\cite{vaitiekenas2020}. In this code the superconductor is integrated out into a self-energy boundary condition. The normal-state Hamiltonian used in the numerical simulations is given by
\begin{align}
	 H_\mathrm{SM}&=\left[(\vec{p}+e \vec{A})^2/(2m_\mathrm{SM})+\Phi(\vec{r}) \right]\sigma_0+\vec{\alpha}\,\cdot\left(\vec{\sigma} \times  (\vec{p}+e \vec{A})\right)+Bg\frac{\mu_B}{2}\sigma_z,\label{eq:Hsm}
\end{align}
where $\sigma_i$ are the Pauli matrices in spin space, $\Phi$ is the potential energy and $\vec{\alpha}$ is the spin-orbit coupling.
We solve for the electrostatic potential in a separate step using the Thomas-Fermi approximation, analogous to what is done in Ref.~\cite{winkler2018}.
In the electrostatic calculations, all surface and dielectric charge densities are set to zero. The vector potential $\vec{A}=\left(0, 0, By\right)$ corresponds to a spatially homogeneous magnetic field in $x$-direction.
In the semiconductor (InSb) we take $m_\mathrm{SM}=0.013\, m_0$, $\vec{\alpha}=(0, 0.1, 0)$\,eV\,nm and $g=-50$~\cite{Vurgaftman2001}.

The superconductor Hamiltonian is given by
\begin{align}
  H_{\rm SC}=\left[ \vec{p}^2/(2 m_\mathrm{SC}) - E_F \right] \tau_z + \Delta(B, \vec{r}) \sigma_y \tau_y,
\end{align}
where $\tau_i$ are the Pauli matrices in Nambu space and the pairing in the superconductor is given by
\begin{equation}
    \Delta(B, \vec{r}) = \Delta_0(\vec{r}) e^{i \varphi(B, \vec{r})} \left(1-\frac{B^2}{B^2_\mathrm{crit}}\right),
\end{equation}
with $\varphi(B, \vec{r})$ the superconducting phase obtained by minimizing the supercurrent as in Ref.~\cite{winkler2018} and $B_\mathrm{crit}=1$\,T is the critical magnetic field of the superconductor. The superconductor is integrated out into a self-energy boundary condition such that the Green's function of the system can be written as~\cite{vaitiekenas2020}
\begin{align}
G^{-1}(\omega) &= \omega - H_\mathrm{SM} - \Sigma_\mathrm{SC}(\omega) .
\label{eq:Green}
\end{align}
In the superconductor (Al) we take $m_\mathrm{super}=m_0$ and $E_F=11.7\,$eV. 

Fig. 3a in the main text shows the energy gap of the simulated island device obtained by calculating the lowest energy poles of Eq.~\eqref{eq:Green}. Fig. 3b in the main text shows the bulk topological phase diagram. For this calculation, we write the Hamiltonian for an infinite length wire as a function of momentum $p_x$ along the wire. The electrostatic potential is taken from a transverse cut through the center of the device. The topological index is calculated using the PFAPACK package~\cite{wimmer2012} and the gap is obtained by the algorithm described in Ref.~\cite{nijholt2016}. 

\bibliography{references}
\clearpage

